\documentclass[aps, prd, twocolumn, floats, floatfix, superscriptaddress, nofootinbib]{revtex4}

\usepackage[utf8]{inputenc}
\usepackage{amsfonts,amsmath,amssymb,amsthm,mathtools}
\usepackage{afterpage}
\usepackage[dvipsnames]{xcolor}
\usepackage[sort&compress]{natbib}
\usepackage{changepage} 

\usepackage{graphicx}
\graphicspath{{figures/}} 

\usepackage{latexsym}
\usepackage{psfrag}
\usepackage{ifthen}
\usepackage[citecolor=blue,colorlinks=true]{hyperref}
\usepackage{float}
\usepackage{lineno}
\usepackage[title]{appendix}
\usepackage{aas_macros}
\usepackage{multirow}
\usepackage{diagbox}
\usepackage{cancel}
\usepackage{siunitx}

\def\assumptionmuchi{\textbf{A.1}}
\def\assumptionM{\textbf{A.2}}
\def\assumptionguai{\textbf{A.3}}

\newcommand{\A}{{\mathcal A}}
\newcommand{\B}{{\mathcal B}}
\newcommand{\C}{{\mathcal C}}
\def\pert{\varepsilon}
\def\manifold{\mathcal{M}}

\def\gback{g}
\def\fpt{K_1}
\def\spt{K_2}

\def\tOmega{\tilde{\Omega}}

\def\xideform{\Xi}

\def\rstar{R}
\def\quotomega{\Delta}

\def\diff{{\rm{d}}}
\def\dim{\text{dim}}

\def\constK{C}

\def\one{\mathrm{a}}
\def\two{\mathrm{b}}
\def\three{\mathrm{c}}

\def\press{\mathcal{P}}

\def\vecb{\begin{pmatrix}n_0\\p_0\end{pmatrix}}
\def\vecbl{\begin{pmatrix}n_0 \eta_\ell\\p_0 \Phi_\ell\end{pmatrix}}

\def\vecpress{\begin{pmatrix}\press_{\ell n}\\\press_{\ell p}\end{pmatrix}}

\def\normal{\mathfrak{n}}

\def\termomega{\mathcal{L}}

\def\AC{\tiny\mbox{AC}}
\def\xiAC{\xi^{\tiny\mbox{AC}}}

\begin{document}

\title{\boldmath A revised formalism for slowly-rotating superfluid neutron stars in general relativity\unboldmath}

\author{Eneko Aranguren}
\email{enekoarangurenruiz@gmail.com}
\affiliation{Department of Physics, University of the Basque Country UPV/EHU, P.O.~Box 644, 48080 Bilbao, Spain}
\author{Jos\'e A.~Font}
\email{j.antonio.font@uv.es}
\affiliation{Departamento de Astronom\'ia y Astrof\'isica, Universitat de Val\`encia, Dr. Moliner 50, 46100, Burjassot (Valencia), Spain}
\affiliation{Observatori Astron\`omic, Universitat de Val\`encia, Catedr\'atico Jos\'e Beltr\'an 2, 46980, Paterna, Spain}
\author{Nicolas Sanchis-Gual}
\email{nicolas.sanchis@uv.es}
\affiliation{Departamento de Astronom\'ia y Astrof\'isica, Universitat de Val\`encia, Dr. Moliner 50, 46100, Burjassot (Valencia), Spain}
 \author{Ra\"ul Vera}
\email{raul.vera@ehu.eus}
\affiliation{Department of Physics, University of the Basque Country UPV/EHU, P.O.~Box 644, 48080 Bilbao, Spain}

\begin{abstract}
We discuss slowly-rotating, general relativistic, superfluid neutron stars in the Hartle-Thorne formulation. The composition of the stars is described by a simple two-fluid model which accounts for superfluid neutrons and all other constituents. We apply a perturbed matching framework to derive a new formalism for slowly-rotating superfluid neutron stars, valid up to second-order perturbation theory, building on the original formulation reported by Andersson and Comer in 2001. The present study constitutes an extension of previous work in the single-fluid case where it was shown that the Hartle-Thorne formalism needs to be amended since it does not provide the correct results when the energy density does not vanish at the surface of the star. We discuss in detail the corrections that need to be applied to the original two-fluid formalism in order to account for non vanishing energy densities at the boundary. In the process, we also find a correction needed in the computation of the deformation of the stellar surface in the original two-fluid model in all cases (irrespective of the value of the energy density at the surface). The discrepancies found between the two formalisms are illustrated by building numerical stellar models, focusing on the comparison in the calculation of the stellar mass, the deformation of the star, and in the Kepler limit of rotation. In particular, using a toy-model equation of state for which the energy density does not vanish at the boundary of the star we demonstrate that the corrections to the formalism we find impact the structure of slowly-rotating superfluid neutron stars in a significant way.
\end{abstract}

\maketitle


\section{Introduction}

In their seminal work in the 1960s, Hartle and Thorne (HT hereafter) formulated the general relativistic treatment of isolated slowly-rotating compact stars in equilibrium~\cite{Hartle1967,Hartle2} {composed of a perfect fluid interior rotating rigidly}. This analytical model provides a perturbative framework to describe the equilibrium configuration of an isolated compact body up to second order in perturbations in General Relativity, around a static, spherically-symmetric configuration. The slow-rotation approximation entails expanding the metric {fields and the matter fields} to $O(\Omega^2)$, where $\Omega$ is the angular velocity of the star. The resulting formalism accurately describes equilibrium models of typical pulsars but it is inappropriate for relativistic stars when rotation approaches the mass-shedding limit~\cite{Weber:1991,Salgado:1994}.

Recent studies  have shown that the mass of a slowly-rotating relativistic star described by a barotropic equation of state (EOS) computed using the HT formalism needs to be amended if {the energy density does not vanish at the surface of the star, since that produces discontinuities of the} second-order perturbation fields there, i.e.~at the matching surface between the interior and exterior solutions for the Euler-Einstein system~\cite{ReinaVera2015,ReinaVeranote,reina-sanchis-vera-font2017,MRV2}. Only for EOS for which the energy density vanishes at the boundary of the star, the expression  provided by~\cite{Hartle1967} to compute the mass of the star yields the correct value. However, there may be situations for which the energy density can exhibit a discontinuity at the stellar surface, notably in the case of EOS describing strange stars (see e.g.~\cite{Colpi:1992}). In particular~\cite{reina-sanchis-vera-font2017} computed equilibrium configurations of the rotating strange stars configurations studied in~\cite{Colpi:1992} finding significantly higher values for the total mass when accounting for the correction to the computation of the mass in the HT formalism. More precisely, the maximum mass found is $\sim 11$\% larger than that attained in the original HT model. In addition, the perturbed-matching approach of~\cite{ReinaVera2015} was also applied in~\cite{reina-sanchis-vera-font2017}  to the tidal problem in binary systems. It was found that it fully accounts for the
correction to the Love numbers needed to obtain truly universal I-Love-Q relations, i.e.~valid for {\it both} neutron stars and strange stars, yielding in a natural way the ad-hoc corrections used in~\cite{Yagi:2013}. 

The equilibrium stellar models built using 
the HT formalism (including those 
in~\cite{reina-sanchis-vera-font2017}) typically 
assume that   relativistic star matter can be described by a single fluid, an over-simplified but widely used premise. However, the composition of a neutron star is far from being made of one single fluid but rather it includes different constituents. The interior of the star has a $\sim 1-2$ km deep solid crust in the outer layers with ions, electrons, nucleii, and a superfluid neutron gas in the inner crust and a $\sim 10$ km core containing superfluid neutrons, superconducting protons, and electrons. Close to the center of the star the composition is almost unknown and proposals involve the possible existence of exotic particles like hyperons, kaon/pion condensates due to phase transitions, or deconfined quarks. A formalism to build slowly-rotating models of superfluid neutron stars in general relativity was presented by Andersson and Comer in~\cite{Andersson_Comer_2001} (AC hereafter) {within the framework of the HT model}. This formalism describes the neutron star matter as a two-fluid model where one of the fluids is the superfluid neutron and the other fluid is a mixture of all other constituents, including protons. As we show here, the amendment of the computation of the mass of the star found in the HT formalism for the single-fluid case is also necessary for the two-fluid model. This has implications in the formalism to build equilibrium configurations of slowly-rotating superfluid neutron stars~\cite{Andersson_Comer_2001} or the computation of observables like quasi-normal modes of oscillation~\cite{Comer1999}. The aim of this work is to present the correction in the HT formalism for the two-fluid model.

In this paper we apply the perturbed matching framework of~\cite{ReinaVera2015,MRV2}
to derive a new formalism for slowly-rotating superfluid neutron stars, valid up to second-order perturbation theory, building on the original two-fluid formulation put forward by AC~\cite{Andersson_Comer_2001}.
{Incidentally, we also find that the computation of the deformation of the star in \cite{Andersson_Comer_2001} by following surfaces of constant energy density needs to be replaced by surfaces of constant ``pressure''. Note that in the single-fluid (perfect) case with barotropic EOS, those two approaches lead to the same result (as so is done in \cite{Hartle1967}), but for the two-fluid model that is not the case.}
We discuss the corrections that need to be applied in order to account for the possibility of discontinuous matter fields at the matching surface {and
the determination of the deformation}.
We first apply our approach to the specific star model considered by~\cite{Andersson_Comer_2001}, highlighting the discrepancies found between the two approaches. Next, using a toy-model EOS for which the energy density does not vanish at the boundary of the star we show that the {amendments} of the AC formalism we {present} in this work must be taken into account as they impact the structure and equilibrium of stationary models of superfluid neutron stars, namely their total mass and radius. This has also consequences for the deformation of the star as well as for the tidal problem,  as it was shown for the single-fluid model previously discussed by~\cite{reina-sanchis-vera-font2017}.

This article is organized as follows: In Sections~\ref{SFNS} and \ref{perturbation-scheme} we discuss the two-fluid model, following the formulation described in \cite{Comer1999}, and our perturbative scheme for the two-fluid case, respectively. These first two sections serve as an introduction to the formalism and lay the foundation for the notation that will be used later. The following sections focus on developing the model to obtain the equations in the background configuration (Section~\ref{background}), both at first order (Section~\ref{first-order}) and at second order (Section~\ref{second-order}). Each of these sections deals with the inner problem, the outer problem, the link between both regions, and the computational procedures carried out to solve both problems. Next, in Section~\ref{sec:deformations} we study the deformation of the star comparing our results with those from \cite{Andersson_Comer_2001} and in Section~\ref{Kepler} we discuss the Kepler or mass-shedding limit of rotation. Numerical results are reported in Section~\ref{numerical} by employing, firstly, 
the same two-fluid EOS as that used by~\cite{Andersson_Comer_2001} and, secondly, a toy-model EOS for which the matter fields are discontinuous at the stellar surface. Finally, our conclusions are presented in Section~\ref{conclusions}.

\section{Superfluid neutron stars}
\label{SFNS}

We start by briefly introducing the formalism of the two-fluid model for superfluid neutron
stars as presented by~\cite{Langlois:1998,Comer1999}. We leave the proper physical motivation and the description of the model to those references and \cite{Andersson_Comer_2001},
and we here focus only on the operational procedures involved.

The central quantity to model general relativistic superfluid neutron stars is the so-called `master' function
\begin{align}
    \Lambda = \Lambda(n^2,p^2,x^2),
\end{align}
which depends on $n^2=-n_\alpha n^\alpha$, $p^2=-p_\alpha p^\alpha$ and $x^2=-p_\alpha n^\alpha$, where $n^\alpha$ and $p^\alpha$ are given by
\begin{align}
    n^{\alpha}=nu^{\alpha},\quad p^{\alpha}=pv^{\alpha},
\end{align}
and $u^{\alpha}$ and $v^{\alpha}$ are the unit timelike vectors describing the flow of protons and neutrons, respectively, and $n$ ($p$) is the neutron (proton) number density.
As stated in  \cite{Comer1999}, the master function encodes the local thermodynamic state of the fluid and serves as a Lagrangian for deriving the superfluid field equations. When taking the limit to a perfect fluid, $\Lambda$ corresponds to minus the total energy density of the fluid.

\subsection{Auxiliary definitions and energy-momentum tensor}

From these quantities~\cite{Comer1999}
define the following set of convenient auxiliary objects.
First, using the short-hand notation for the first derivatives of $\Lambda$ with respect to its three arguments 
\begin{align*}
&\A:=-\frac{\partial\Lambda(n^2,p^2,x^2)}{\partial x^2},\quad
\B:=-2\frac{\partial\Lambda(n^2,p^2,x^2)}{\partial n^2},\\
&\C:=-2\frac{\partial\Lambda(n^2,p^2,x^2)}{\partial p^2},
\end{align*}
the following one-forms are defined
\begin{equation*}
    \mu_{\alpha}:=\B n_{\alpha}+\A p_{\alpha},\qquad\chi_{\alpha}:=\C p_{\alpha}+\A n_{\alpha},
\end{equation*}
which are dynamically and thermodynamically conjugate to $n^\alpha$ and $p^\alpha$,
respectively~\cite{Comer1999}.
After defining the generalized pressure as
\begin{equation}\label{eq:generalizedpressure}
    \Psi:=\Lambda-n^{\alpha}\mu_{\alpha}-p^{\alpha}\chi_{\alpha},
\end{equation}
the energy-momentum tensor of the fluid is then given by
\begin{equation}\label{eq:Tmunu}
  T^\alpha{}_{\beta}=\Psi \delta^\alpha_{\beta}+p^\alpha \chi_\beta+n^\alpha \mu_\beta.
\end{equation}

\subsection{Equations of the fluid}

The equations of motion are given by the number density conservation equations
\begin{equation}
  \nabla_\alpha n^\alpha=0,\quad \nabla_\alpha p^\alpha=0,\label{eq:conservation}
\end{equation}
and by the Euler equations
\begin{equation}\label{eq:euler}
  n^\alpha (\nabla_{\alpha} \mu_{\beta}-\nabla_{\beta} \mu_{\alpha})=0,\quad
    p^\alpha (\nabla_{\alpha} \chi_{\beta}-\nabla_{\beta} \chi_{\alpha})=0.
\end{equation}
Equations \eqref{eq:conservation} and \eqref{eq:euler} imply $\nabla^\alpha T_{\alpha\beta}=0$.

\subsection{Stationary and axisymmetric configurations, circularity condition
and rigid rotation}

We next assume that the spacetime and the fluid are stationary and axisymmetric,
and that the flows satisfy the circularity condition,
i.e. they rotate around the axis (there are no convective motions),
and rotate rigidly.
In consequence, following the conventions and notation in \cite{Andersson_Comer_2001}, we take a
coordinate system $\{t,r,\theta,\phi\}$ adapted to the symmetries, 
so that $\partial_t$ is a timelike Killing vector field and
$\partial_\phi$ is an axial Killing vector field (with regular axis),
the pair $\{\theta,\phi\}$ coordinates the sphere, and the metric has the form
\begin{align}\label{g_stax}
  g_{\tiny{\mbox{STAX}}}=&-\left(N^2-\sin^2\theta K (N^\phi)^2\right) \diff t^2
   +V\diff r^2\\
  &-2\sin^2\theta K N^\phi \diff t \diff \phi + K\left(\diff\theta^2+\sin^2\theta \diff\phi^2\right),\nonumber
\end{align}
where all functions depend only on $r$ and $\theta$,
and the vectors $u$ and $v$ (we use index-free notation when convenient) satisfy
\begin{align}\label{eq:uv}
  u=\frac{1}{\sqrt{N^2-\sin^2\theta K (N^\phi-\tOmega_n)^2}}
  \left(\partial_t+\tOmega_n \partial_\phi\right),\nonumber\\
  v=\frac{1}{\sqrt{N^2-\sin^2\theta K (N^\phi-\tOmega_p)^2}}
  \left(\partial_t+\tOmega_p \partial_\phi\right),
\end{align}
for some constants $\tilde\Omega_n$ and $\tilde\Omega_p$, which represent the angular velocities of neutrons and protons, respectively.

Eqs.~\eqref{eq:conservation} are automatically satisfied, and
Eqs.~\eqref{eq:euler} are equivalent to \cite{Andersson_Comer_2001}
\begin{equation}\label{eq:euler_const}
  \mu_c=-g(\partial_t+\tOmega_n\partial_\phi,\mu),
  \quad
  \chi_c=-g(\partial_t+\tOmega_p\partial_\phi,\chi),
\end{equation}
for some constants $\mu_c$ and $\chi_c$.
Note we use $g(\cdot,\cdot)$ for the scalar product in the index-free notation.

The functions (of $r$ and $\theta$)
involved in the two-fluid problem are thus given by the list
$\{N,N^\phi,K,V\}$ plus the four constants $\{\tOmega_p,\tOmega_n,\mu_c,\chi_c\}$.

\subsection{Global configuration: vacuum exterior and surface of the star}
\label{sec:surface_star}

The global configuration for an isolated  finite star is built by the
matching of two spacetimes with boundary,
one to describe the interior region $(\manifold^+,g^+)$ with boundary $\Sigma^+$,
that solve the two-fluid model just described,
and a vacuum exterior region $(\manifold^-,g^-)$ with boundary $\Sigma^-$.
The matching procedure ensures the point-to-point
identification of the two boundaries to form the so-called
matching hypersurface  $\Sigma\equiv \Sigma^+=\Sigma^-$.
The interior and exterior problems consist of the corresponding equations,
with ``regularity'' conditions at the origin and at infinity, plus some relations of the boundary data on $\Sigma^\pm$ provided by the matching conditions (so that there are no energy surface layers at the boundary). The matching conditions determine also the form of $\Sigma$, and thus provide the surface of the star. In this section we obtain the equation that determines the surface of the star in the two-fluid model. In the following we use the $+$ and $-$ indexes to refer to interior and exterior quantities, respectively.

Given that the exterior region is vacuum, the matching conditions imply
\begin{equation}
   Ein(g^+)_{\alpha\beta}\normal_+^\alpha|_{\Sigma_+}=0,\label{eq:Israel_full}
\end{equation}
where $\normal_+^\alpha$ is normal to the hypersurface $\Sigma^+$,
and we use $Ein(g)$ to denote the Einstein tensor computed from $g$.
The vector $\normal_+^\alpha$,
defined in principle only on  $\Sigma^+$, is an unknown of the problem.

The assumption that the whole configuration is stationary and axisymmetric,
that is, that both the interior and exterior regions are stationary and axisymmetric
and that the boundaries $\Sigma^+$ and $\Sigma^-$
preserve those symmetries (see \cite{Vera2002}),
implies that $\normal_+^\alpha$ must be orthogonal to the Killings $\partial_t$ and $\partial_\phi$.
This implies, in particular, that $\normal_+^\alpha$ is spacelike.
It is then chosen to point from $\manifold^+$ inwards, and
normalised to one.
Analogously, we will have another normal  $\normal_-^\alpha$
defined on $\Sigma^-$ to point $\manifold^-$ outwards,
to be identified to $\normal_+^\alpha$ at each point of $\Sigma$.

For the two-fluid model above, for which in particular Eq.~\eqref{eq:uv} holds, we have necessarily that the vectors $u^\alpha$ and $v^\alpha$ (on $\Sigma^+$)
are orthogonal to $\normal_+^\alpha$, and hence
also are $n^\alpha$, $p^\alpha$, $\mu^\alpha$ and $\chi^\alpha$ by construction.
Therefore, given \eqref{eq:Tmunu},
condition \eqref{eq:Israel_full} is written equivalently as
\[
  (\Psi \normal_+^\alpha+ p^\alpha \chi_\beta \normal_+^\beta+ n^\alpha \mu_\beta \normal_+^\beta)|_{\Sigma^+}=
  \Psi \normal_+^\alpha|_{\Sigma^+}=0,
\]
and thus
\begin{equation}
  \label{eq:Psi_S_full}
  \Psi|_{\Sigma^+}=0.
\end{equation}
In terms of the chart $\{t,r,\theta,\phi\}$ introduced above,
this equation can be writen as
\[\Psi(r,\theta)=0,\]
which is the equation that defines $\Sigma^+$ in an implicit manner
in terms of $r$ and $\theta$.

Condition \eqref{eq:Psi_S_full} is a necessary condition for the matching, but it is not sufficient. Nevertheless, as shown in~\cite{Vera:2003} (see also \cite{Mars:1998})
that condition will be the only one involving only the interior side. The rest of the matching conditions provide the matching hypersurface
from the other side $\Sigma^-$ and relations between the boundary data for the interior and exterior problems.

\section{Perturbation scheme to second order}
\label{perturbation-scheme}

In this section we introduce the ingredients
and procedures we will follow in subsequent sections to set and solve the stationary and axisymmetric perturbative model of the two-fluid star around static and spherically symmetric background configuration. We follow the stationary and axisymmetric perturbative scheme to second order around a static and spherically symmetric background $(\manifold,g)$ as described in \cite{ReinaVera2015} (see also \cite{MRV2})
based in an abstract perturbation parameter $\pert$. We refer to \cite{MRV1} for the set of definitions involved in a perturbation scheme,
which basically consists of a family of spacetimes $(\manifold_\pert,\hat g_\pert)$
  with $(\manifold_0,\hat g_0)=(\manifold,g)$
together with a class of point identification (gauges). On the other hand, we will try to stick close to the notation  in \cite{Andersson_Comer_2001}
for the names of the functions relative to the
background, first and second order perturbations.

\subsection{The geometry}

We are thus given a static and spherically symmetric background spacetime $(\manifold,\gback)$
with
\begin{equation}\label{eq:gback}
  \gback=-e^{\nu(r)}\diff t^2+e^{\lambda(r)}\diff r^2+r^2(\diff \theta^2+\sin^2\theta \diff \phi^2),
\end{equation}
and first and second-order perturbation tensors on $(\manifold,\gback)$
\begin{align}
\fpt  =&
 -2r^2\, \omega(r) \sin ^2 \theta \diff t \diff \phi \label{fopert_tensor},\\
\spt = &\left(-4 e^{\nu(r)} h(r, \theta) + 2 r^2{\omega}^2(r) \sin^2\theta\right)\diff t^2 \label{sopert_tensor}\\
& + 4e^{\lambda(r)} v(r, \theta) \diff r^2 +4 r^2 k(r, \theta)
(\diff \theta^2 + \sin ^2 \theta \diff \phi^2).\nonumber
\end{align}
The family of tensors relative to the perturbation scheme
to second order in terms of a (so far abstract) parameter $\pert$,
$g_\pert=\gback+\pert\fpt+\frac{\pert^2}{2}\spt+O(\pert^3)$,
corresponds to the form \eqref{g_stax} with the correspondences
\begin{align}
  &N=e^{\nu(r)/2}\left(1+\pert^2 h(r,\theta)\right),\nonumber\\
  &V=e^{\lambda(r)}\left(1+2\pert^2 v(r,\theta)\right),\nonumber\\
  &K=r^2\left(1+2\pert^2 k(r,\theta)\right),\quad N^\phi=\pert\omega(r,\theta).\label{NKV}
\end{align}

\subsection{The two-fluid}

The two-fluid interior is assumed to be based on some function $\Lambda$ with three arguments.
The form of $\Lambda$ is given by the background configuration, and it is assumed
that such form is kept in the perturbative scheme (see \cite{MRV1}).
This is analogous to the single-fluid case (the HT model) in which the same barotropic EOS is imposed at all the perturbative levels, in particular at the background. Explicitly, one demands that $\Lambda_\pert(n_\pert^2,p_\pert^2,x_\pert^2)=\Lambda(n^2_\pert,p_\pert^2,x_\pert^2)$, where $n_\pert$, $p_\pert$, and $x_\pert$ will correspond
to the fluid functions for each value of $\pert$.

Following \cite{Andersson_Comer_2001} we  use the expansions
\begin{align}\label{eq:nepe}
  &n_\pert(r,\theta)=n_0(r)(1+\pert^2 \eta(r,\theta)),\nonumber\\
  &p_\pert(r,\theta)= p_0(r)(1+\pert^2\Phi(r,\theta)),
\end{align}
and the notation $\Lambda_\pert:=\Lambda(n_\pert^2,p_\pert^2,x_\pert^2)$,
so that $\Lambda_0=\Lambda(n_0^2,p_0^2,x_0^2)$.
More explicitly we will also use
$\Lambda_0(r):=\Lambda(n_0^2(r),p_0^2(r),x_0^2(r))$
and equivalently for $\Psi_0(r).$

The perturbation of the velocity vector of the fluids
is assumed to be driven by the rotations
\begin{align}
  \tOmega_n=\pert\Omega_n,\quad
  \tOmega_p=\pert\Omega_p,\label{eq:Omegas_epsilon}
\end{align}
for some constants $\Omega_n$ and $\Omega_p$,
so that the rotation enters at first order of $\pert$
(and not at second order, see \cite{MRV2}).
Introducing the correspondences \eqref{NKV} into \eqref{eq:uv}
we have $u_\pert=u_{(0)}+\pert u_{(1)}+\frac{1}{2}\pert^2 u_{(2)}+O(\pert^3)$
and $v_\pert=v_{(0)}+\pert v_{(1)}+\frac{1}{2}\pert^2 v_{(2)}+O(\pert^3)$ with
\begin{align*}
  u_{(0)}=&e^{-\nu(r)/2}\partial_t,\quad
  u_{(1)}=e^{-\nu(r)/2}\Omega_n\partial_\phi,\\
  u_{(2)}=&\left(-2e^{-\nu(r)/2}h(r,\theta)\right.\\
  &\left.+e^{-3\nu(r)/2}r^2\sin^2\theta(\omega(r)-\Omega_n)\right)\partial_t,\\
   v_{(0)}=&e^{-\nu(r)/2}\partial_t,\quad
   v_{(1)}=e^{-\nu(r)/2}\Omega_p\partial_\phi,\\
  v_{(2)}=&\left(-2e^{-\nu(r)/2}h(r,\theta)\right.\\
  &\left.+e^{-3\nu(r)/2}r^2\sin^2\theta(\omega(r)-\Omega_p)\right)\partial_t.
\end{align*}
Now, from $n_\pert^\alpha=n_\pert u_\pert^\alpha$ and $p_\pert^\alpha=p_\pert v_\pert^\alpha$
we have
\begin{align*}
  &n_\pert^\alpha=n_0\left\{\!u_{(0)}^\alpha+\pert u_{(1)}^\alpha+\frac{\pert^2}{2}(u_{(2)}^\alpha+2\eta u_{(0)}^\alpha)\!\right\}+O(\pert^3),\\
  &p_\pert^\alpha=p_0\left\{v_{(0)}^\alpha+\pert v_{(1)}^\alpha+\frac{\pert^2}{2}(v_{(2)}^\alpha+2\Phi v_{(0)}^\alpha)\right\}+O(\pert^3),
\end{align*}
from where, using the definition $x_\pert^2=-p_\pert{}_\rho n_\pert^\rho$,
we obtain
\begin{align}\label{eq:xe}
  x_\pert^2(r,\theta)=&n_0(r) p_0(r)\left\{ 1+\frac{1}{2}\pert^2\big(2\Phi(r,\theta)+2\eta(r,\theta)\right.\nonumber\\
  &\left.+e^{-\nu(r)}r^2\sin^2\theta (\Omega_n-\Omega_p)^2\big) \right\}+O(\pert^3).
\end{align}
Then, in particular,
\begin{equation}\label{eq:x0}
  x_0^2(r)=n_0(r) p_0(r).
\end{equation}

The Einstein field equations must hold now for each member of the family
of the perturbation scheme parametrised with $\pert$, that is
\begin{equation}\label{eq:Einstein_e}
  Ein(g_\pert)^\alpha{}_\beta=\varkappa T_\pert{}^\alpha{}_\beta,
\end{equation}
where $\varkappa=8\pi G/c^4$ and with the obvious extensions of the previously defined quantities
to the $\pert$-family given by
\begin{align*}
  &T_\pert{}^\alpha{}_\beta=\Psi_\pert \delta^\alpha_{\beta}+p_\pert^\alpha \chi_\pert{}_\beta+n_\pert^\alpha \mu_\pert{}_\beta,\\
  &\Psi_\pert=\Lambda_\pert-n_\pert^{\rho}\mu_\pert{}_{\rho}-p_\pert^{\rho}\chi_\pert{}_{\rho},\\
  &\mu_\pert{}_\alpha=\B_\pert n_\pert{}_{\alpha}+\A_\pert p_\pert{}_{\alpha},\quad
    \chi_\pert{}_{\alpha}=\C_\pert p_\pert{}_{\alpha}+\A_\pert n_\pert{}_{\alpha},\\
  &\A_\pert=-\frac{\partial\Lambda(n_\pert^2,p_\pert^2,x_\pert^2)}{\partial x_\pert^2},\quad\B_\pert=-2\frac{\partial\Lambda(n_\pert^2,p_\pert^2,x_\pert^2)}{\partial n_\pert^2},\\
  &\C_\pert=-2\frac{\partial\Lambda(n_\pert^2,p_\pert^2,x_\pert^2)}{\partial p_\pert^2}.
\end{align*}
The explicit dependence on $r$ and $\theta$
of these quantities is obtained
once we introduce the expressions of the corresponding functions
$n_\pert$ and $p_\pert$ from \eqref{eq:nepe}
and $x_\pert$ from \eqref{eq:xe}. In the following
we will use the notation
$\mathcal{O}_0:=\mathcal{O}_\pert|_{\pert=0}$ for any object $\mathcal{O}$.
Observe that
$\A_0$, $\B_0$ and $\C_0$ are functions that depend only on $r$.

As mentioned above,
the only remaining equations we have to impose
are
the Eulerian equations for the fluid  \eqref{eq:euler_const}. These equations
applied to each member of the family of the perturbation scheme
just read
\begin{align}\label{eq:euler_e}
  &\mu_c{}_\pert=-g_\pert(\partial_t+\pert\Omega_n\partial_\phi,\mu_\pert),
  \nonumber\\
  &\chi_c{}_\pert=-g_\pert(\partial_t+\pert\Omega_p\partial_\phi,\chi_\pert).
\end{align}
The expansions on $\pert$ of the left hand sides are taken to
be of the form
\begin{align}
  \label{eq:const_euler_2}
  \mu_c{}_\pert=\mu_\infty\left(1+\pert^2 \gamma_n\right)+O(\pert^3),\nonumber\\
  \chi_c{}_\pert=\chi_\infty\left(1+\pert^2 \gamma_p\right)+O(\pert^3),
\end{align}
that define
the four constants $\mu_\infty(=\mu_c{}_0)$, $\chi_\infty(=\chi_c{}_0)$, $\gamma_n$ and $\gamma_p$.

\subsubsection{The perturbation parameter $\pert$}
\label{sec:pert}

The introduction of the perturbation parameter $\pert$ by means
of \eqref{eq:Omegas_epsilon} has been performed,
as mentioned earlier, to incorporate the rotation at first (and only at first)
order. The fact that this can be done in the perfect fluid case
has been argued many times in the literature 
(see e.g. \cite{Hartle1967}), and has been finally shown in full in \cite{MRV2}.
In doing so we have passed from two parameters in the ``full exact'' case,
$\tOmega_n$ and $\tOmega_p$ to three, namely $\pert$, $\Omega_n$ and $\Omega_p$.
The scalability property of perturbative schemes introduces the freedom of redefining the perturbation parameter, and that freedom
has translated here into the introduction of the spurious information
provided by one extra parameter. The perturbation scheme does not depend on that choice
(let us refer to  \cite{MRV2} for a full account on the subject in the perfect fluid case).

From a computational point of view the procedure consists on choosing freely the value of one paramenter amongst $\Omega_p$, $\Omega_n$ and $\pert$, solve the problems, and then simply use the scalability property to fix the model to the data we want. A particular choice can consist on setting $\pert=1$.
Instead, following \cite{Andersson_Comer_2001} in essence, we will use the parameter
\begin{equation}
  \label{def:Delta}
  \quotomega=\frac{\tOmega_n}{\tOmega_p},
\end{equation}
which equals, by construction,  $\quotomega=\Omega_n/\Omega_p$,
and eventually fix $\pert=\tOmega_p$, so that  $\Omega_p=1$
and the perturbation depends on
two parameters, $\quotomega(=\Omega_n)$ and $\pert=\tOmega_p$.
With this choice $\pert$ contains the numerical value of the (rigid) rotation of
the fluid of protons.

This choice of parametrisation will be imposed only when needed,
when we eventually describe the computational process. The reason is
for convenience, since
we prefer to leave all the construction 
in terms of $\Omega_n$ and $\Omega_p$ and keep all expressions
symmetric with respect to the two fluids.
Motivated by the notation used in some parts of \cite{Andersson_Comer_2001},
we will use a ``little hat" notation to indicate that some quantity $f$ has been calculated by setting $\Omega_n=\Delta$, $\Omega_p=1$ in the corresponding equations
to obtain $\hat f$. This means that, for any quantity $f_1$ at first order we have $\hat f_1=f_1/\Omega_p$ by definition, while for any quantity $f_2$ at second order,  $\hat f_2=f_2/\Omega_p^2$. Observe that $\pert f_1=\tOmega_p \hat f_1$
and $\pert^2 f_2=\tOmega^2_p \hat f_2$ because $\pert=\tOmega_p/\Omega_p$.

Finally, let us note that given \eqref{eq:Omegas_epsilon}, the parameters
$\tOmega_p$ and $\tOmega_n$ here correspond to
$\Omega_p$ and $\Omega_n$ used in \cite{Andersson_Comer_2001}.

\subsection{The global perturbation scheme}
\label{pert_scheme}
The background configuration follows the construction
as described in Section \ref{sec:surface_star}, and it is thus
divided into the interior part $(\manifold^+,\gback^+)$ with boundary $\Sigma^+$
and the exterior part $(\manifold^-,\gback^-)$ with boundary $\Sigma^-$,
with $\gback^\pm$ of the form \eqref{eq:gback}
and identified boundaries as $\Sigma\equiv \Sigma^+=\Sigma^-$.
Although we ought to be using $\{t_+,r_+,\theta_+,\phi_+\}$
and  $\{t_-,r_-,\theta_-,\phi_-\}$
for the charts at the exterior and interior domains respectively,
we are going to use a common name $\{t,r,\theta,\phi\}$
to simplify the notation whenever that does not lead to confusion.

For a global static and spherically symmetric background configuration,
the matching in the background configuration
is performed assuming that the spherical symmetry and staticity
of the exterior and interior regions is preserved by the matching \cite{Vera2002}.
As a result, the boundaries $\Sigma^\pm$
are defined in each case by $r_+=\rstar_+$ and $r_-=\rstar_-$, respectively,
for some pair of positive reals $\rstar_+$ and $\rstar_-$,
and we take $r_+\in(0,\rstar_+)$ for the interior and $r_-\in(\rstar_-,\infty)$
for the exterior. {The normal vectors are then given by $\normal^\pm_0=-e^{-\lambda_\pm(\rstar_\pm)/2} \partial_r^\pm$, that are to be identified by the matching procedure.}
We will then use the usual notation $[f]:=f_+|_{r_+=\rstar_+}-f_-|_{r_-=\rstar_-}$,
where $f_+$ and $f_-$ are functions defined on $\manifold^+$ and $\manifold^-$ respectively.

From the perturbation scheme one constructs the set of
interior and exterior problems for the families of metrics $\gback_\pert^+$
and $\gback_\pert^-$. In the interior
the equations at first and second order
correspond to the first and second derivatives of \eqref{eq:Einstein_e}
for $g^+_\pert$
with respect of $\pert$ evaluated at $\pert=0$, respectively.
The equations on the exterior are
obtained equivalently, but now using \eqref{eq:Einstein_e}
for $g^-_\pert$ with the right hand side set to zero.
At first order the functions involved will be those appearing
in $\fpt$, and at second order, those in $\spt$ plus those in $\fpt$ feeding
the ``inhomogeneous'' part of the equations. If we call $f^+$ and $f^-$
the set of functions involved at some order on each region, the procedure
thus provides equations on $(\manifold^+,g^+)$ and $(\manifold^-,g^-)$ for $f^+$ and $f^-$.

The matching procedure using the perturbation scheme provides, firstly, relations between the boundary data of the functions $f$ on $\Sigma^+$ and $\Sigma^-$ by means of their differences, or jumps, $[f]$, and possibly $[f']$.
The equations that determine these jumps
will depend, in general, on the two classes of (spacetime) gauges used to construct the interior
and exterior problems (one gauge at the interior and another at the exterior), and also on the class of gauges involved in the construction
of the family of matching hypersurfaces 
$\Sigma_\pert\equiv\Sigma_\pert^+=\Sigma_\pert^-$ as subsets in
the corresponding $\manifold^\pm$ (see \cite{ReinaVera2015} for an expanded description),
which we refer to as the hypersurface gauge \cite{Mars:2005,Mars:2007}.

Secondly, the matching procedure provides
the deformation of the matching hypersurface,
that is encoded in the family of hypersurfaces
$\Sigma^\pm_\pert$ (one family at each side).
That deformation is described at each point
on the respective $\Sigma^\pm=\Sigma^\pm_0$ by a vector
$Z^\pm_1$ at first order and
$Z^\pm_2$ at second order.
At the interior, $Z^+_1$ corresponds to the velocity vector at points on $\Sigma^+$
of the curves that follow the points on the family $\Sigma^+_\pert$
identified by the hypersurface gauge followed by the spacetime
gauge at the $+$ side,
and $Z^+_2$ corresponds to the acceleration of that curve on $\Sigma^+$.
The same goes for the exterior region with $-$.
Each of the four vectors $Z^\pm_{1/2}$ can be decomposed as tangent $T^\pm_{1/2}$
and normal parts to the corresponding $\Sigma^\pm$ as
$Z^\pm_{1}=T^\pm_{1}+ Q^\pm_{1}\normal^\pm_0$ and $Z^\pm_{2}=T^\pm_{2}+ Q^\pm_{2}\normal^\pm_0$,
where $Q_1^\pm$ and $Q_2^\pm$ are two pairs of functions defined on their respective
$\Sigma^\pm$.
The deformation of $\Sigma$ described by $\Sigma_\pert$ as a set of points
corresponds then, at each side, to the normal part of  $Z^\pm_{1/2}$, that is
$Q_1^\pm$ at first order, and $Q_2^\pm$ at second order. The deformations refer to the (spacetime) gauges used at each side.
To sum up,
the perturbed matching conditions to second order will provide relations between
\begin{itemize}
\item[(i)]  the jumps of the functions, and possibly derivatives,
  of the background configuration,
\item[(ii)]the jumps $[f]$, and possibly $[f']$, of the functions $f$ involved
  in $\fpt$ and $\spt$
\item[(iii)]the differences of the tangent vectors $T^\pm_1$ and $T^\pm_2$,
\item[(iv)]two pairs of functions
  $Q_1^\pm$ and $Q_2^\pm$ (at each correponding side)
  that describe the deformation of the surface of the star (as seen from each side and with respect to the gauge used).
\end{itemize}
Let us stress that these relations follow from a pure geometrical
setting.
The perturbed matching between the two sides $\pm$ will exist if there exist
functions $Q_1^\pm$ and $Q_2^\pm$ and vectors  $T^\pm_1$ and $T^\pm_2$
such that all the perturbed matching conditions
are satisfied. On the other hand,
in principle, those conditions may not provide a closed system for all the jumps
and  $Q_1^\pm$ and $Q_2^\pm$,
that is, that some freedom can be left (and some not even fixed by a choice of gauge,
see~\cite{Mars:2007}).

Nevertheless, when the field equations are imposed, possibly
complemented with some other matter-field conditions at the boundary (as for instance
no layer of electrical charge),
more conditions on the jumps of the relevant functions may appear.
One expects that the system of equations
for the jumps and the functions $Q_1^\pm$ and $Q_2^\pm$
in terms of background quantities
closes, and that the solution exists.

Finally, let us remark that the quantities involved in this
perturbative scheme are, in general, gauge dependent. Part
of the analysis of the problem consists of the control over all those
dependencies.
In particular, $Q_2^\pm$ are both spacetime and hypersurface-gauge
dependent. However, in the case
of stationary and axisymmetric perturbations around static and spherically symmetric
backgrounds, the deformation of the hypersurface (at second order)
appears in the matching equations
in terms of alternative functions $\xideform^\pm$
constructed from $Q_2^\pm$, as shown in \cite{ReinaVera2015,MRV2},
  that are hypersurface-gauge invariant if $Q_1=0$
  and also invariant under the class of gauges at second order that maintain
  the form of the second order perturbation tensor used in the more general analysis there.
  In the present case, that is, for the class of gauges in which
  $\fpt$ and $\spt$ have the form \eqref{fopert_tensor} and \eqref{sopert_tensor},
  we have
  \[
    \xideform^\pm=Q^\pm_2+\kappa(T^\pm_1,T^\pm_1).
  \]
  A hypersurface gauge (see  \cite{ReinaVera2015,MRV2})
  can always be chosen so that
  either $T_1^+$ or $T_1^-$ vanishes (but not both). Choosing, e.g., $T_1^-=0$
  we are left with $\xideform^-=Q^-_2$.

Next we deal with the building of the background
configuration, and follow with the first and the second order problems.
In each case we present the equations of the interior region, the solution
of the vacuum exterior, and the equations that provide the matching
of the two problems, that will be taken from
the geometrical analysis in \cite{MRV2} (see also \cite{ReinaVera2015})\footnote{
  The perturbed matching (and thus the quantities $Q_1^\pm$ and $Q_2^\pm$ and
  the vector fields $T_1^\pm$ and $T_2^\pm$)
  is assumed in  \cite{ReinaVera2015} to be axially symmetric.
  The general analysis is made in \cite{MRV2}, where it is shown 
  that axial symmetry of the perturbed matching is a necessary consequence of the whole setting.}.
Each section will end with an account
on the explicit procedure used to solve the corresponding global interior-exterior
problem.

\section{The background configuration}
\label{background}

\subsection{Interior problem}
Once the function $\Lambda$ is set 
the list of quantities that describe the background configuration
of the interior of the star is given by
$\{\lambda_+(r),\nu_+(r),n_0(r),p_0(r)\}$ plus the pair
of constants $\{\mu_\infty,\chi_\infty\}$.
The equations are found from \eqref{eq:Einstein_e} and
\eqref{eq:euler_e} evaluated at $\pert=0$.
From Eq.~\eqref{eq:Einstein_e} we obtain
\begin{align}
&\lambda_+^{\prime}=+\frac{1-e^{\lambda_+}}{r}-\varkappa re^{\lambda_+}\Lambda_0,\label{eq:dlamb}\\
&\nu_+^{\prime}=-\frac{1-e^{\lambda_+}}{r}+\varkappa re^{\lambda_+}\Psi_0,\label{eq:dnu}
\end{align}
while Eq.~\eqref{eq:euler_e} yields \cite{Comer1999}
\begin{align}
  &\mu_\infty e^{-\nu_+(r)/2}=\B_0(r) n_0(r)+\A_0(r) p_0(r)=:\mu_0(r),\label{eq:euler_back1} \\
  &\chi_\infty e^{-\nu_+(r)/2}=\A_0(r) n_0(r)+\C_0(r) p_0(r)=:\chi_0(r),\label{eq:euler_back2}
\end{align}
where we take advantage of the expressions to introduce two auxiliary
functions $\mu_0$ and $\chi_0$.\footnote{Then, for the one-form
  $\mu_\pert{}_\alpha$ we have
  $\mu_{0}{}_\alpha \diff x^\alpha=\mu_0(r)e^{\nu(r)/2}\diff t$.}
It is important to stress that if $\mu_0$ vanishes
  at some point then it vanishes everywhere, and the same for $\chi_0$.
The radial derivatives of the two equations in \eqref{eq:euler_back1}-\eqref{eq:euler_back2} provide, respectively,
\begin{align}
&\A^0_0 p_0^{\prime}+\B^0_0 n_0^{\prime}+\frac{1}{2}(\B_0n_0+\A_0p_0)\nu_+^{\prime}=0,\label{eq:first1}\\
&\C^0_0 p_0^{\prime}+\A^0_0 n_0^{\prime}+\frac{1}{2}(\A_0n_0+\C_0p_0)\nu_+^{\prime}=0,\label{eq:first2}
\end{align}
with\footnote{Our $\A^0_0$, etc., correspond to $\A^0_0|_0$, etc., in \cite{Andersson_Comer_2001}.}
\begin{align*}
  \A^0_0:=&\A_0+2\frac{\partial\B_0}{\partial p_0^2}n_0p_0
    +2\frac{\partial\A_0}{\partial n_0^2}n_0^2
    +2\frac{\partial\A_0}{\partial p_0^2}p_0^2\\
    &+\frac{\partial\A_0}{\partial x_0^2}n_0p_0,\\
  \B^0_0:=&\B_0+2\frac{\partial\B_0}{\partial n_0^2}n_0^2
    +4\frac{\partial\A_0}{\partial n_0^2}n_0p_0
    +\frac{\partial\A_0}{\partial x_0^2}p_0^2,\\
  \C^0_0:=&\C_0+2\frac{\partial\C_0}{\partial p_0^2}p_0^2
    +4\frac{\partial\A_0}{\partial p_0^2}n_0p_0
    +\frac{\partial\A_0}{\partial x_0^2}n_0^2.
\end{align*}
Observe that with our notation the following chain of identities
(and the analogous) hold
\[\frac{\partial\B_0}{\partial p_0^2}
  =\frac{\partial\B(n_0^2,p_0^2,x_0^2)}{\partial p_0^2}
  =\left.\frac{\partial\B(n_\pert^2,p_\pert^2,x_\pert^2)}{\partial p_\pert^2}\right|_{\pert=0}.
  \]

For convenience, let us define the auxiliary functions (for each region $+$ and $-$) $j(r)$ and $M(r)$ by
\begin{align*}
    j(r):=e^{-(\lambda(r)+\nu(r))/2},\quad
    1-\frac{2M(r)}{r}:=e^{-\lambda(r)}.
\end{align*}
The function $M(r)$ corresponds to the (Misner-Sharp) mass,
and \eqref{eq:dlamb} can be re-expressed as \cite{Andersson_Comer_2001}
\[
  M(r)=4\pi\int_0^r s^2(-\Lambda_0(s))ds.
\]

We end this subsection with some remarks.
First, the invertibility of the system \eqref{eq:first1}-\eqref{eq:first2} must be kept under control. For that
we can re-express \eqref{eq:first1}-\eqref{eq:first2} as
\begin{equation*}
  \mathfrak{A}\vecb'=-\frac{1}{2}\nu'_+\mathfrak{B}\vecb
\end{equation*}
with
\begin{equation}
  \label{eq:matrices}
  \mathfrak{A}:=
  \begin{pmatrix}
    \B^0_0 & \A_0^0\\
    \A_0^0 & \C^0_0
  \end{pmatrix},\qquad
  \mathfrak{B}:=
  \begin{pmatrix}
    \B_0 & \A_0\\
    \A_0 & \C_0
  \end{pmatrix}.
\end{equation}
Therefore $\Lambda$ will have to satisfy the condition that $\mathfrak{A}$
is invertible in all the range $r\in(0,\rstar_+)$,
so that the system \eqref{eq:first1}-\eqref{eq:first2} is equivalent to
\begin{equation}
  \label{eq:first_matrix}
  \vecb'=-\frac{1}{2}\nu'_+ \mathfrak{A}^{-1}\mathfrak{B}\vecb.
\end{equation}
It must be kept in mind, however, that we had
\eqref{eq:euler_back1}-\eqref{eq:euler_back2}, which reads
\[
  \mathfrak{B}(r)\vecb(r)=\begin{pmatrix}\mu_0\\\chi_0\end{pmatrix}(r)
  =e^{-\nu_+(r)/2}\begin{pmatrix}\mu_\infty\\\chi_\infty\end{pmatrix},
\]
and therefore \eqref{eq:first_matrix} can be also writen as
\begin{equation}
  \label{eq:first_matrix_good}
  \vecb'
  =-\frac{1}{2}\nu'_+\mathfrak{A}^{-1}e^{-\nu_+/2}\begin{pmatrix}\mu_\infty\\\chi_\infty\end{pmatrix}.
\end{equation}
The importance of this form of the equations for $n_0(r)$ and $p_0(r)$ is that
it provides $n_0'$ and $p_0'$ at all points where $\mathfrak{A}$ is invertible,
even at points where $n_0$ and $p_0$ vanish, where $\mathfrak{B}$ necessarily diverges.

Also, we have by construction the equalities, c.f. \eqref{eq:euler_back1}-\eqref{eq:euler_back2},
\begin{align}\label{eq:E+P}
  \Psi_0(r)-\Lambda_0(r)=&n_0(r)\mu_0(r)+p_0(r) \chi_0(r)\\
  =&e^{-\nu_+(r)/2}(\mu_\infty n_0(r)+\chi_\infty p_0(r)),\nonumber
\end{align}
and also
\begin{equation}
  \label{eq:e+p_geom}
  \Psi_0(r)-\Lambda_0(r)=-\frac{1}{\varkappa}\frac{(j_+^2(r))^{\prime}e^{\nu_+(r)}}{r}
\end{equation}
by \eqref{eq:dlamb} and \eqref{eq:dnu} in terms of the metric functions.

A straightforward calculation using the chain rule with
\eqref{eq:first_matrix} and \eqref{eq:E+P} allows us to write
\begin{equation}
  \label{eq:psi'}
  \Psi'_0(r)=-\frac{1}{2}\nu'(r)(\Psi_0(r)-\Lambda_0(r)),
\end{equation}
and (we avoid the $r$ dependence and the $+$ ``interior'' subindex)
  \begin{align}
    \label{eq:Lambda'}
    \Lambda_0'=&\frac{1}{2}\nu'\frac{1}{\det \mathfrak{A}}
    \left(\C^0_0\mu_0^2-2\mu_0\chi_0\A^0_0+\B^0_0\chi_0^2\right)\\
    =&\frac{1}{2}\nu'e^{-\nu}\frac{1}{\det \mathfrak{A}}
    \left(\C^0_0\mu_\infty^2-2\mu_\infty\chi_\infty\A^0_0+\B^0_0\chi_\infty^2\right)\nonumber
  \end{align}
  after using \eqref{eq:first_matrix_good} in the last equality.

  Differentiating equation \eqref{eq:dnu}, and using
    \eqref{eq:psi'} together with \eqref{eq:dlamb}-\eqref{eq:dnu},
    we get an equation for $\nu^{\prime\prime}$ only in terms of $\lambda$
\begin{equation}
2r\nu^{\prime\prime}_++\nu_+^{\prime}(r\nu_+^{\prime}-2)-\lambda_+^{\prime}(2+r\nu_+^{\prime})+\frac{4}{r}(e^{\lambda_+}-1)=0,\label{eq:ddnu}
\end{equation}
which is the same equation that in the perfect fluid case
arises because of the isotropy of the pressure.

The fact that the metric $g$ is smooth at the origin
implies that $\lambda(r)$ and $\nu(r)$ are smooth up to the boundary
and admit the expansions (see a full proof, e.g.,  in \cite{MRV2})
\[
  \lambda(r)=\lambda_0+\lambda_2r^2+O(r^4),\quad
  \nu(r)=\nu_0+\nu_2r^2+O(r^4).
\]
Introducing this in the field equations \eqref{eq:dlamb} and \eqref{eq:dnu}
we obtain
\begin{align}
  &\lambda(0)=\lambda_0=0,\quad
  \lambda_2=-\frac{1}{3}\varkappa\Lambda_0(0),\nonumber\\
  &\nu_2=\frac{1}{2}\varkappa\left(\Psi_0(0)-\frac{1}{3}\Lambda_0(0)\right).
  \label{eq:l2v2}
\end{align}
Observe that $\nu_0$ remains free, and that accounts to the freedom of
shifting the ``Newtonian'' potential.
Asking that the potential is zero at infinity
in the global problem will fix that value, see below.
It is also worth noticing that, \emph{a priori}, the functions
$n_0(r)$ and $p_0(r)$ need not follow the same pattern
as the metric functions around the origin (expansion in even powers of $r$).
However, taking
\begin{align*}
  &n_0(r)=n_0(0)+n_{0,1}r+n_{0,2}r^2+O(r^3),\\
  &p_0(r)=p_0(0)+p_{0,1}r+p_{0,2}r^2+O(r^3),
\end{align*}
the Eulerian equations \eqref{eq:first_matrix} evaluated on $r=0$,
since $\nu'(0)=0$,
imply that $n_{0,1}=p_{0,1}=0$ in particular.

Let us now stress the fact that the existence of the solution
to the perturbative scheme for perfect fluids
requires that the sum of the central energy density and pressure
does not vanish at the origin \cite{MRV2}. Although
the full problem has not been dealt with rigour,
later we will find that in the two-fluid model we will need
to ask, equivalently, that
$\Psi_0(0)-\Lambda_0(0)\neq 0$.
As a result, c.f. \eqref{eq:E+P}, we shall
demand that
\begin{itemize}
\item[\assumptionmuchi]$\mu_\infty$ and $\chi_\infty$ cannot both be zero.
\end{itemize}

The assumption $\Psi_0(0)-\Lambda_0(0)\neq 0$ can be motivated
  on physical grounds
  by requiring the positivity of the mass $M(r)$ around the origin,
  and thus that $-\Lambda_0(0)>0$, plus the positivity of the effective pressure.

\subsection{Exterior solution}

The outern vacuum region $(\manifold^-,\gback^-)$ is defined
by $\Lambda=0$, $n_0=0$, $p_0=0$, and is thus described by the set
$\{\lambda_-(r),\nu_-(r)\}$.
The solution of field equations is
\begin{equation}\label{eq:back_sol_ext}
  e^{-\lambda_{-}(r)}=e^{\nu_{-}(r)}=1-\frac{2M}{r}, 
\end{equation}
for some constant $M$,
that hold over some domain $r\in(\rstar_-,\infty)$
for some $\rstar_->0$.
Observe that the solution can equivalently
  be charaterised by $M_-(r)=M$ and $j_-(r)=1$ for  $r\in(\rstar_-,\infty)$.
We assume in the following that
\begin{itemize}
\item[\assumptionM] $M>0$
\end{itemize}
so that we
trivially recover Schwarzschild with mass $M$
on $r\in(\rstar_-,\infty)$ with the ``usual'' coordinates.

\subsection{Matching of the problems}

The matching of the background configuration is given by the conditions
(see e.g. \cite{MRV2})
\begin{align}\label{eq:match_back}
    \rstar_-=\rstar_+=:\rstar,\quad[\nu]=0,\quad [\nu']=0,\quad[\lambda]=0.
\end{align}
Using that the exterior solution is given by \eqref{eq:back_sol_ext}, these
conditions imply
\begin{align}
  e^{\lambda_{+}(\rstar)}=e^{-\nu_{+}(\rstar)}=\frac{\rstar}{\rstar-2M},
  \label{eq:bg_boundary}
\end{align}
and
\begin{equation}
  \label{eq:nup_boundary}
  \nu_+'(\rstar)=\frac{1}{\rstar}\frac{2M}{\rstar-2M}.
\end{equation}
In terms of $M$ and $j$ we have
$$M=M_-(\rstar),\qquad j_-(\rstar)=1.$$
Introducing these relations on \eqref{eq:dnu} we obtain
\begin{equation}
  [\Psi_0]=0.\label{eq:P_S}
\end{equation}
This is the only consequence the matching conditions have on the functions
describing the matter content.

Note $[\Lambda_0]=\Lambda_0{}(\rstar):=\Lambda(n^2_0(\rstar),p_0^2(\rstar),x^2_0(\rstar))$
and equivalently
$[\Psi_0]=\Psi_0{}(\rstar)$ because the corresponding functions on the exterior vanish
identically.
From \eqref{eq:E+P} we can write $\Psi_0(\rstar)=0$ equivalently as
\begin{align}
  \label{eq:Lambda_S}
  \Lambda_0(\rstar)=&-e^{-\nu_+(\rstar)/2}(\mu_\infty n_0(\rstar)+\chi_\infty p_0(\rstar))\nonumber\\
  =&-\sqrt{\frac{\rstar}{\rstar-2M}}(\mu_\infty n_0(\rstar)+\chi_\infty p_0(\rstar)),
\end{align}
after using
\eqref{eq:bg_boundary} in the last equality.
Equation \eqref{eq:P_S} shows that \emph{the matching hypersurface (the value of $\rstar$ in
  this case) is determined by the vanishing of $\Psi_0$}, i.e. the first solution
of $\Psi_0(\rstar)=0$,
whereas $\Lambda_0(\rstar)$ attains the value given by \eqref{eq:Lambda_S}. That value will depend upon the explicit
form of $\Lambda$ in terms of its three arguments.
In the perfect fluid case that corresponds to the equation of state,
see e.g. \cite{ReinaVera2015,MRV2}.

A rigorous treatment of the existence and uniqueness
  problem (that will be presented elsewhere)
  will also need an assumption on the behaviour of $\Lambda_0$
  at the boundary, namely that if it vanishes then both
  $n_0$ and $p_0$ must also vanish, that is,
  \begin{itemize}
  \item[\assumptionguai] $\Lambda_0(\rstar)=\Lambda(n^2_0(\rstar),p_0^2(\rstar),x^2_0(\rstar))=0 \implies n_0(\rstar)=p_0(\rstar)=0$.
  \end{itemize}
  Let us stress that if one demands $\mu_\infty=\chi_\infty$, as in the models in \cite{Andersson_Comer_2001}, assumption \assumptionmuchi{} as well as \assumptionguai{}
  (because of \eqref{eq:Lambda_S}) are automatically satisfied.

The first and second order matching conditions
  will contain jumps of higher  derivatives of $\lambda$ and $\nu$. For later use,
  then, we present next the  expressions relating those differences with the
fluid quantities on the surface. Using \eqref{eq:match_back} taking into account
the field equations \eqref{eq:dlamb} and \eqref{eq:ddnu} (and yet another radial derivative)
we obtain
\begin{align}
    &[\lambda^{\prime}]=-\varkappa \rstar e^{\lambda(\rstar)} [\Lambda_0],\label{eq:dlambda_S}\\
    &[\lambda'']=-\rstar e^{\lambda(\rstar)}\varkappa[\Lambda_0']+[\lambda'^2],\label{eq:ddlambda_S}\\
    &[\nu^{\prime\prime}]=-\varkappa\left(1+\frac{\rstar\nu^{\prime}(\rstar)}{2}\right)e^{\lambda(\rstar)}[\Lambda_0],\label{eq:ddnu_S}\\
    &[\nu''']=\frac{1}{\rstar}\left(1+\frac{\rstar \nu'(\rstar)}{2}\right)[\lambda'']+C_\nu[\lambda'],\label{eq:dddnu_S}
\end{align}
where $C_\nu$ is a constant whose explicit form will not be needed,
that relate the geometical jumps with the jumps of the matter field.

In the following, for any quantity $f$
satisfying  $[f]=0$ we will use simply $f(\rstar)=f_+(\rstar)=f_-(\rstar)$
and just $f$ if $f$ is only defined on the boundary.

\subsection{Solving the background global problem}
\label{sec:background_comp}

To sum up,  given a function of three arguments $\Lambda$,
the background interior is described by four functions
$\{\lambda_+(r),\nu_+(r),n_0(r),p_0(r)\}$ that satisfy the equations
\eqref{eq:dlamb}, \eqref{eq:dnu}, \eqref{eq:first_matrix}.
on the domain $r\in(0,\rstar)$ for some constant $\rstar>0$
that is fixed by the matching procedure below.
Two first integrals to the system are given by \eqref{eq:euler_back1}-\eqref{eq:euler_back2}.
Moreover, it must be stressed that
$\nu_+(r)$ only enters the equations algebraically through $\nu_+'(r)$.
As a result, the system of equations provide $\nu_+'(r)$, and thus $\nu_+(r)$
up to a free additive constant.
This is equivalent to the fact
that $\nu_0:=\nu_+(0)$ is left undetermined
by the interior problem, and has to be fixed \emph{a posteriori} by imposing
that $\nu_+(\rstar)=-\lambda_+(\rstar)$ (c.f. \eqref{eq:match_back}) as follows.

The procedure is to integrate the equations \eqref{eq:dlamb}, \eqref{eq:dnu},
 \eqref{eq:first_matrix}, replacing $\nu(r)$ by some function $\tilde\nu(r)$,
from the origin (at $r=0$)
given the conditions \eqref{eq:l2v2}
plus finite values for $n_0(0)$ and $p_0(0)$, together with the
condition $\tilde\nu(0)=0$.
The interior problem thus integrated, which
is independent of $\tilde\nu(0)$, provides
$\lambda_+(r)$, $n_0(r)$ and $p_0(r)$, and therefore also  $\Psi_0(r)$.
Because of the matching condition \eqref{eq:P_S}, and given that the exterior
is vacuum, $\rstar$ is obtained as the (first) zero $\Psi_0(\rstar)=0$.
Now we just have to set $\nu_+(r)=\tilde\nu(r)-\tilde\nu(\rstar)-\lambda_+(\rstar)$
to have the complete solution for the interior.

The global solution is completed once the exterior
parameter $M$ is obtained from \eqref{eq:bg_boundary}.

\section{First order problem}
\label{first-order}

The equations are found by differentiating once
\eqref{eq:euler_e} and \eqref{eq:Einstein_e}
with respect to $\pert$ and evaluating at $\pert=0$. From  \eqref{eq:euler_e} we do not obtain anything, while the field equations
provide the equations for $\omega_+(r)$ and $\omega_-(r)$ on their respective regions.
The matching will be provided by Proposition 1 in \cite{ReinaVera2015} (see also \cite{MRV2}).

\subsection{Equations at the interior}

On the interior region  the equation for $\omega_+(r)$ is given by
\begin{equation}
\frac{1}{r^3}\left(r^4j \omega_+^{\prime}\right)^{\prime}=2j\termomega,\label{eq:domega}
\end{equation}
where here we use the convenient definition
\begin{align*}
    \termomega:=\varkappa r e^{\lambda_+}\left(L_n n_0 \mu_0+L_p p_0 \chi_0\right)
\end{align*}
with \cite{Andersson_Comer_2001}
\begin{equation*}
  L_n:=\omega_+-\Omega_n,\quad L_p:=\omega_+-\Omega_p.
\end{equation*}
Later we will use the equality
\begin{align}\label{eq:L_auxiliar}
  \termomega=
  &\varkappa re^{\lambda_+} L_n(\Psi_0-\Lambda_0)+\varkappa re^{\lambda_+} (\Omega_n-\Omega_p)\chi_0 p_0\nonumber\\
  =&L_n(\nu_+'+\lambda_+')+\varkappa re^{\lambda_+}(\Omega_n-\Omega_p)\chi_0 p_0,
\end{align}
that follows from \eqref{eq:E+P}, \eqref{eq:dlamb} and \eqref{eq:dnu}.

\subsection{Exterior solution}

The equation for $\omega_-(r)$ is \eqref{eq:domega} with $\termomega=0$. The solution that vanishes
at infinity is
\begin{equation}\label{eq:omega_exterior}
    \omega_{-}(r)=\frac{2J}{r^{3}},
\end{equation}
for some constant $J$. Observe that we can always choose this exterior solution,
vanishing at infinity, fixing the first order gauge
on the exterior region \cite{ReinaVera2015,MRV2}.

\subsection{First order matching}

The first order matching is provided by Proposition B.1 in \cite{MRV2},
  see also Proposition 1 in \cite{ReinaVera2015}.
Let us recall that the result in \cite{MRV2} generalises that in \cite{ReinaVera2015}
in that $Q_1^\pm$ and the vectors $T_1^\pm$
can depend on all the coordinates on the boundary
$\{\tau,\vartheta,\varphi\}$,
and the  condition $\nu'(\rstar)\neq 0$
was missing in the statement in Proposition 1 in \cite{ReinaVera2015}.
Observe first that the conditions $\nu'(\rstar)\neq 0$ and
$2e^{\lambda(\rstar)}-2+\rstar \nu'(\rstar)\neq 0$ are satisfied because $M\neq 0$.
By fixing the first order gauge on the interior region
(using the gauge vector $V_1=b_1t\partial_\phi$, see Proposition 2.5. in \cite{MRV2})
we can also choose $b_1=0$ in Proposition B.1 in \cite{MRV2})
so that the interior and exterior
problems are matched (in those fixed first order gauges) by
\begin{equation}
  [\omega]=0,\quad  [\omega^{\prime}]=0,\label{eq:match_omega}
\end{equation}
and the deformation quantities $Q_1^\pm(\tau,\vartheta,\varphi)$ satisfy
\begin{equation}\label{eq:Q1_match_geo}
  [Q_1]=0,\quad Q_1[\lambda']=0,\quad Q_1[\nu'']=0.
\end{equation}
Using \eqref{eq:dlambda_S} and \eqref{eq:ddnu_S} the above three conditions turn into
\begin{equation}
  \label{eq:Q1_match}
  [Q_1]=0,\quad Q_1\Lambda_0(\rstar)=0.
\end{equation}
Consequently, given assumption \assumptionguai{} we also have
\begin{equation}
  \label{eq:Q1_match_}
  Q_1 p_0(\rstar)=0,\quad Q_1 n_0(\rstar)=0.
\end{equation}
It is convenient, for later use, to use these matching conditions
together with the equation \eqref{eq:domega} on each region
to obtain
\begin{align}\label{eq:match_omegapp}
  &[\omega^{\prime\prime}]=\frac{2}{\rstar}\termomega(\rstar) +\frac{1}{2}(\lambda'_+(\rstar)+\nu_+'(\rstar))\omega^{\prime}(\rstar)\nonumber\\
  &=\varkappa e^{\lambda(\rstar)}\left\{-\Lambda_0(\rstar)\!\left(2(\omega(\rstar)-\Omega_n)+\frac{1}{2}\rstar\,\omega^{\prime}(\rstar)\!\right)\right.\nonumber\\
  &\,\,\left.+2(\Omega_n-\Omega_p)\chi_0(\rstar)p_0(\rstar)\frac{}{}\right\}.
\end{align}
The above properties of $Q_1$ thus imply
\[
  Q_1[\omega'']=0.
\]
  
\subsection{Solving the first order global problem}
To be consistent with \cite{Andersson_Comer_2001} we follow initially
the procedure used there
to calculate the first order global problem for the functions $\omega_+$
and $\omega_-$.
We already have the exterior
solution \eqref{eq:omega_exterior}. As for the interior, instead of
computing $\omega_+$, we use
$\hat L_n(r):=\omega_+(r)/\Omega_p-\Omega_n/\Omega_p$,
and use the quotient $\quotomega=\Omega_n/\Omega_p$ as the free parameter
of the problem together with $\tOmega_p$.

Equation \eqref{eq:domega} is equivalent to
\begin{align}\label{eq:ode_L}
  &\frac{1}{r^4}\left(r^4j(r) \hat L'_n(r)\right)^{\prime}-2\varkappa e^{(\lambda_+(r)-\nu_+(r))/2}(\Psi_0-\Lambda_0)\hat L_n\nonumber\\
  &=2\varkappa e^{(\lambda_+(r)-\nu_+(r))/2}(\quotomega-1)\chi_0(r) p_0(r)
\end{align}
after making use of \eqref{eq:L_auxiliar}. This equation
  corresponds to (44) in \cite{Andersson_Comer_2001} before dividing by $\Omega_p$.
Given \eqref{eq:omega_exterior}, the matching conditions \eqref{eq:match_omega}
imply (compare with (62) and (63) in \cite{Andersson_Comer_2001})
\begin{align}
  \hat L_n(\rstar)=-\quotomega+ \frac{1}{\Omega_p} \frac{2J}{R^3},\quad
  \hat L'_n(\rstar)=-\frac{1}{\Omega_p}\frac{6 J}{R^4}\nonumber\\
  \Longrightarrow\quad \hat L_n'(\rstar)=-\frac{3}{R}\left(\hat L_n(\rstar)+\quotomega\right).
  \label{eq:Ln_surface}
\end{align}
The pole structure of \eqref{eq:ode_L}
under the assumption that $\Psi_0(0)-\Lambda_0(0)$ does not vanish
implies that the homogeneous part of \eqref{eq:ode_L} admits a
unique bounded solution up to a scaling factor (see e.g. Lemma D.2 in \cite{MRV2}).
Therefore, operationaly, to integrate the interior problem
we start with a function $G(r)$ that satisfies the homogeneous part
of \eqref{eq:ode_L} (the left hand side) and assume  $G(0)=1$.
Next, we take a particular solution $F(r)$ of the full equation \eqref{eq:ode_L}
integrating from the origin under the assumption $F(0)=0$. It can be shown
that such particular solution, vanishing at the origin, is unique.
Then
\begin{equation*}
  \hat L_n(r)=\alpha G(r)+ F(r),
\end{equation*}
with
\begin{equation*}
  \alpha=-\frac{3(F(\rstar)+\quotomega)+\rstar F'(\rstar)}{3 G(\rstar)+\rstar G'(\rstar)}
\end{equation*}
solves \eqref{eq:ode_L} and satisfies  \eqref{eq:Ln_surface}.
The global solution for $\omega$ is thus given by
\begin{align}
  &\omega_+(r)=\Omega_p\left(\alpha G(r)+ F(r)+\quotomega\right),\quad \omega_-(r)=\frac{2J}{r^3}
  \nonumber\\
  &\mbox{with}  \label{eq:sol_omega} \\
  &J=-\frac{1}{6}\rstar^4\omega'_+(\rstar)=-\frac{1}{6}\Omega_p\rstar^4\left(\alpha G'(\rstar)+F'(\rstar)\right).
  \nonumber
\end{align}
Using equation \eqref{eq:domega} one can rewrite the
value of $J$ as an integral of $\termomega$, and thus recover
the integral expression for $J$
found in equation (59) in \cite{Andersson_Comer_2001}.

{As explained in Subsection \ref{sec:pert},
we use the ``little hat" notation
$\hat f:= f/\Omega_p$ for any first order quantity $f$,
in particular $\hat\omega_\pm:=\omega_\pm/\Omega_p$.}

\subsubsection{Rotation and angular momentum}

Let us recall that the rotation of the fluids are given by
\eqref{eq:Omegas_epsilon}, and
$\tOmega_n$ and $\tOmega_p$ are taken to be the two parameters
of the model, that we take to be in the form of $\Delta$ and $\tOmega_p$.
Also, the cross term in $\diff t\diff\phi$ at first
order in the family of metrics $g_\pert$
is given by $\pert\omega(r)$ with \eqref{eq:sol_omega}.
As a result, the total angular momentum $J^S$ of the solution
(to first order), and thus of the star configuration, is given by $J^S=\pert J=\tOmega_p \hat J$,
and therefore
\begin{equation}
  J^S=-\frac{1}{6}\tOmega_p\rstar^4\left(\alpha G'(\rstar)+F'(\rstar)\right).
  \label{eq:J}
\end{equation}

\section{Second order problem}
\label{second-order}

The problem at second order is tackled using an expansion in Legendre polynomials
$P_\ell(\cos\theta)$. We follow \cite{Andersson_Comer_2001}
and take the usual form for the functions $h(r,\theta)$, $v(r,\theta)$ and
$k(r,\theta)$ at both domains (the interior $\manifold^+$ and the exterior $\manifold^-$),
given by
\begin{align}
    &h^\pm(r,\theta) = h^\pm_0(r)+h^\pm_2(r)P_2(\cos\theta),\nonumber\\
    &v^\pm(r,\theta) = v^\pm_0(r)+v^\pm_2(r)P_2(\cos\theta),\label{eq:decomp2}\\
    &k^\pm(r,\theta) = k^\pm_2(r)P_2(\cos\theta),\nonumber
\end{align}
where $P_2(\cos\theta)=(3\cos^2\theta-1)/2$, and also
\begin{align*}
    &\eta(r,\theta) = \eta_0(r)+\eta_2(r)P_2(\cos\theta),\\
    &\Phi(r,\theta) = \Phi_0(r)+\Phi_2(r)P_2(\cos\theta),
\end{align*}
at the interior. The second order problem eventually
separates onto a problem
involving only $\ell=0$ functions
and another for $\ell=2$ functions (see below).
This form of the functions entering the second order perturbation tensor
  is assumed in
  \cite{Andersson_Comer_2001} for the two-fluid problem following the arguments
  used in the original Hartle-Thorne model (for a perfect fluid).
  Let us stress here that the fact
  that there exist gauges at the interior and exterior regions
  such that the perturbation tensors to second order for
  any stationary and axisymmetric perturbation for a perfect fluid have this form
  has been shown only recently, and with no need of equatorial symmetry,
  in the two works \cite{MRV1} and \cite{MRV2}.
  Here we just follow \cite{Andersson_Comer_2001} and take this form
  of the second order perturbation functions by assumption. 

In order to have more compact expressions,
and to ease the comparison with the expressions found in \cite{ReinaVera2015},
we introduce some convenient auxiliary definitions related with the second order pressure
that shall be used to substitute the set $\{\eta_\ell(r),\Phi_\ell(r)\}$.
Since $\mathfrak{A}$ is invertible, for each $\ell=0,2$ we define the set of (four) functions
$\{\press_{\ell n}(r),\press_{\ell p}(r)\}$ such that
\begin{align}\label{eq:def_pressl}
  \vecbl=&\mathfrak{A}^{-1}\vecpress\\
  &-(1-\ell)\frac{r^2}{3}e^{-\nu}(\Omega_n-\Omega_p)^2\mathfrak{A}^{-1} \begin{pmatrix}p_0 \mathcal{D}_n\\n_0 \mathcal{D}_p\end{pmatrix}\nonumber
\end{align}
holds, with
\begin{align}
  &\mathcal{D}_n:=\A_0+2n^2_0\frac{\partial\A_0}{\partial n_0^2}
  +n_0 p_0\frac{\partial\A_0}{\partial x_0^2},\nonumber\\
  &\mathcal{D}_p:=\A_0+2p^2_0\frac{\partial\A_0}{\partial p_0^2}
  +n_0 p_0\frac{\partial\A_0}{\partial x_0^2}.\label{def:DnDp}
\end{align}
It is also convenient to define a quantity that depends on $r$ and is constructed
from the background and first order solutions,
plus the parameters $\Omega_n$ and $\Omega_p$, and can be therefore already computed. That is
\begin{align}
     f_{\omega}:=&\frac{1}{6}e^{-(\lambda+\nu)}r^4{\omega^{\prime}}^{2}\nonumber\\
    &+\varkappa\frac{r^{4}}{3e^{\nu}}\left(L_n^2n_0^2\B_0+L_p^2p_0^2\C_0+2L_nL_pn_0 p_0\A_0\right)\nonumber\\
  =&\frac{1}{6}e^{-(\lambda+\nu)}r^4{\omega^{\prime}}^{2}+\varkappa\frac{r^{4}}{3e^{\nu}}\big\{(\Psi_0-\Lambda_0)L_n^2\nonumber\\
  &-\chi_0p_0(L^2_n-L^2_p)-n_0 p_0 \A_0(\Omega_n-\Omega_p)^2\big\}.\nonumber 
\end{align}
Furthermore, let us finally introduce the convenient auxiliary quantities
\begin{align}
  \press_\ell:=&n_0\press_{\ell n}+p_0\press_{\ell p},\label{def:Pl} \\
  \Upsilon_0:=&\eta_0n_0^2\B_0 + \Phi_0p_0^2\C_0+(\eta_0+\Phi_0)x_0^2\A_0\nonumber\\
  =&-\frac{2}{\nu_+'}\left\{\left(\press_{0 n}-\frac{r^2}{3}e^{-\nu}(\Omega_n-\Omega_p)^2\mathcal{D}_n p_0\right)n_0'\right.\nonumber\\
  &\left.+\left(\press_{0 p}-\frac{r^2}{3}e^{-\nu}(\Omega_n-\Omega_p)^2\mathcal{D}_p n_0\right)p_0'\right\},\nonumber
\end{align}

\subsection{Equations at the interior}

The equations are found by differentiating twice
\eqref{eq:Einstein_e} and
\eqref{eq:euler_e}
with respect to $\pert$ and evaluating at $\pert=0$.
We do not specify the $+$ superindex in this subsection.
The Euler equations \eqref{eq:euler_e} with \eqref{eq:const_euler_2},
using \eqref{eq:euler_back1}-\eqref{eq:euler_back2}, provide
\begin{align}
  &\press_{0n}-\mu_0\left\{\frac{r^2}{3}e^{-\nu}L_n^2-h_0+\gamma_n\right\}=0,\nonumber\\
  &\press_{0p}-\chi_0\left\{\frac{r^2}{3}e^{-\nu}L_p^2-h_0+\gamma_p\right\}=0,\label{eq:gammas}\\
  &\press_{2n} + \mu_0\left\{\frac{r^2}{3}e^{-\nu}L_n^2 + h_2\right\} =0,\nonumber\\
&\press_{2p} + \chi_0\left\{\frac{r^2}{3}e^{-\nu}L_p^2 + h_2\right\} =0.\label{eq:euler_l2}
\end{align}
From the field equations \eqref{eq:Einstein_e} we obtain
\begin{align}
   &(r e^{-\lambda} v_0)'=\frac{\varkappa}{2} r^2\left\{\Upsilon_0+\frac{r^2}{3}e^{-\nu}n_0p_0\A_0(\Omega_n-\Omega_p)^2\right\}\nonumber\\
   &+ f_\omega-\frac{1}{12}r^4 e^{-(\lambda+\nu)} \omega'^2,\label{eq:dm0}\\
   &h_0^{\prime}-\left(\nu^{\prime}+\frac{1}{r}\right)v_0\label{eq:dh0}\\
   &= \frac{\varkappa}{2} re^{\lambda}
    \press_{0}
    -\frac{r^3}{12e^{\nu}}\left(2\varkappa e^{\lambda}n_0p_0\A_0\left(\Omega_n-\Omega_p\right)^2+{\omega^{\prime}}^2\right),\nonumber
\end{align}
and
\begin{align}
  &v_2= f_{\omega}-h_2,\label{eq:v2h2}\\
  &(k_2+h_2)^{\prime} = -\nu^{\prime}h_2+f_{\omega}\left(\frac{1}{r}+\frac{\nu^{\prime}}{2}\right),\label{eq:dk2dh2}\\
    &h_2^{\prime}=-\left\{\nu^{\prime}+\frac{1}{r\nu^{\prime}}\left(\frac{2}{r}(e^\lambda-1)-\lambda'-\nu'\right)\right\}h_2\label{eq:dh2}\\
    &-4\frac{h_2+k_2}{r^{2}\nu^{\prime}}e^{\lambda}-\frac{r^2e^{-\nu}}{3\nu^{\prime}}{\omega^{\prime}}^2+\frac{1}{2r^2}\left(\frac{2}{\nu^{\prime}}e^{\lambda}+r^2\nu^{\prime}\right)f_\omega.\nonumber
\end{align}

The comparison with the equations in \cite{Andersson_Comer_2001} is given by the following.
Equations \{\eqref{eq:gammas},\eqref{eq:euler_l2},\eqref{eq:dm0},\eqref{eq:dh0},\eqref{eq:v2h2}\} correspond to \{(42),(43),(47),(49),(45) in \cite{Andersson_Comer_2001}\} one by one respectively, while the set \{\eqref{eq:v2h2},\eqref{eq:dk2dh2},\eqref{eq:dh2}\} is equivalent to
\{(45),(46),(50) in  \cite{Andersson_Comer_2001}\}. The remaining equation (48) in
\cite{Andersson_Comer_2001} can be shown to be a consequence of the rest.

\subsection{Exterior solution}

The functions in the exterior region $\{h_\ell,v_\ell,k_2\}$ satisfy
equations \eqref{eq:dm0}-\eqref{eq:dh2} with vanishing $n_0$, $p_0$
(and thus vanishing $\Upsilon_0$ and $\press_0$) and $\omega=\omega_-$ given
by \eqref{eq:omega_exterior}. The solutions are given for $r\in(R,\infty)$
by \cite{Andersson_Comer_2001,ReinaVera2015}
\begin{align}
    &h_0^{-}(r)=-v^-_0(r)=-\frac{\delta M}{r-2M}+\frac{J^2}{r^3(r-2M)},\label{eq:v0_h0_ext}
\end{align}
and
\begin{align}
    &h_2^{-}(r)=-\constK\bigg\{\frac{3}{2}\frac{r^2}{M^2}
    \left(1-\frac{2M}{r}\right)
    \log\left(1-\frac{2M}{r}\right)\label{eq:h2_ext}\\
    &+\frac{(r-M)(3-6M/r-2(M/r)^2)}{M(1-2M/r)}\bigg\}+\frac{J^2}{Mr^3}\left(1+\frac{M}{r}\right),\nonumber\\
    &k_2^{-}(r)=\constK \bigg\{\frac{3}{2}\frac{r^2}{M^2}
    \left(1-\frac{2M^2}{r^2}\right)
    \log\left(1-\frac{2M}{r}\right)\label{eq:k2_ext}\\
    &+\frac{3(r-M)-8(M/r)^2(r-M/2)}{M(1-2M/r)}\bigg\}-\frac{J^2}{Mr^3}\left(1+\frac{2M}{r}\right),\nonumber\\
    &v_2^{-}(r)=-h_2^{-}(r)+\frac{2J^2}{Mr^3}\left(1-\frac{2M}{r}\right),\label{eq:v2_ext}
\end{align}
where $\delta M$ and $\constK$ are constants.
Let us note that, as in the first order case,
  the gauge at the exterior is fixed so that the solution
vanishes at infinity (see \cite{MRV2}).

\subsection{Second order matching}
For the matching of the second order problem we use
Proposition B.7 in \cite{MRV2} (see also Proposition 2 in \cite{ReinaVera2015},
bearing in mind that Proposition B.7 allows \emph{a priori}
  for an arbitrary deformation of the surface,
  while in \cite{ReinaVera2015} the deformation is assumed to be axially symmetric).
Let us stress that this is a geometrical result, that is, independent of the field equations. We apply Proposition B.7 in \cite{MRV2} for $\mathcal{R}=r$,
  $\mathcal{W}=0$ and $\normal=-e^{\lambda(\rstar)/2}\partial_r$,
noting that the present class of gauges
correspond to $k^\pm_0(r)=f^\pm(r,\theta)=0$ as they appear in \cite{MRV2} and
that $m$ in \cite{MRV2} is $v$ here.
First, given that $h,m,k$ satisfy \eqref{eq:decomp2}
we get $c_0=c_1=H_1=0$.
Then, after using 
the background matching \eqref{eq:match_back} and \eqref{eq:Q1_match_geo},
(so that $Q_1[\omega'']=0$ and the relations involving $(Q_1)^2$ after
(5.68) in \cite{MRV2} hold), we obtain the set of relations
\begin{align}
  &[\xideform]=0,\label{match:Q2}\\
  &[k]=0,\quad [h]=\frac{1}{2}H_0,\nonumber\\
  &[v]-\rstar[k']=\frac{e^{-\lambda(\rstar)}}{4}\left\{e^{\lambda(\rstar)/2}\xideform[\lambda']-(Q_1)^2[\lambda'']\right\}, \label{match:funs}\\
  &[h']-\frac{\rstar}{2}\nu'(\rstar)[k']=\frac{e^{-\lambda(\rstar)}}{4}\left\{e^{\lambda(\rstar)/2}\xideform[\nu'']
  -(Q_1)^2[\nu''']\right\}. \label{match:funs_p}
\end{align}
So far we have not used the field equations, neither at the background level nor at first and second order. This result is, so far, purely geometric.
As explained in \cite{ReinaVera2015,MRV2} (see Proposition 2.5 in \cite{MRV2}),
a change of gauge driven by the vector $V^+_2=H_0t\partial_t$ in the interior region
only affects $h_0^+$ and allows us to set $H_0=0$ in the matching without
loss of generality (but keeping in mind this change has been already used).
This corresponds with a trivial shift of the function $h$ at the interior,
and this is, in turn, a consequence of the usual freedom in the shift
of the Newtonian potential (and $\nu$ in the background configuration).
This choice leaves us with no freedom left in the
spacetime gauges.

If we introduce the background field equations through the
relations \eqref{eq:dlambda_S}-\eqref{eq:dddnu_S}
and write down explicitly the above relations in terms of the functions
introduced in the decompositions \eqref{eq:decomp2}
plus
\begin{align*}
&(Q_1)^2(t,\theta,\phi)=\sum_{\ell=0}^2 \mathcal{Q}_{\ell}(t,\phi)P_\ell(\cos\theta)+\mathcal{Q}_{\perp}(t,\theta,\phi),\\
  &\xideform(t,\theta,\phi)=\sum_{\ell=0}^2\xideform_{\ell}(t,\phi)P_\ell(\cos\theta)+\xideform_{\perp}(t,\theta,\phi),
\end{align*}
the relations
\eqref{match:funs}-\eqref{match:funs_p}
are equivalent to
\begin{widetext}
\begin{align}
   &[h_0]=0,\quad [v_0]=-\frac{1}{4} \rstar e^{\lambda(\rstar)/2}\xideform_{0}\varkappa\Lambda_0(\rstar)+\frac{1}{4}\rstar\mathcal{Q}_{0}\varkappa[\Lambda_0'],\label{match:h0v0}\\
  &[h'_0]=-\frac{1}{4}e^{\lambda(\rstar)/2}\xideform_{0}\left(1+\frac{\rstar\nu^{\prime}(\rstar)}{2}\right)\varkappa\Lambda_0(\rstar)+\frac{1}{4}\mathcal{Q}_{0}\left(1+\frac{\rstar \nu'(\rstar)}{2}\right)\varkappa[\Lambda_0'],\label{match:h0p}\\
  &[k_2]=0,\quad [h_2]=0,\quad [v_2]-\rstar[k_2']=-\frac{1}{4} \rstar e^{\lambda(\rstar)/2}\xideform_{2}\varkappa\Lambda_0(\rstar)+\frac{1}{4}\rstar\mathcal{Q}_{2}\varkappa[\Lambda'_0],\label{match:k2h2v2} \\
   &[h'_2]-\frac{\rstar}{2}\nu'(\rstar)[k'_2]=-\frac{1}{4}e^{\lambda(\rstar)/2}\xideform_{2}\left(1+\frac{\rstar\nu^{\prime}(\rstar)}{2}\right)\varkappa\Lambda_0(\rstar)
     +\frac{1}{4}\xideform_{2}\left(1+\frac{\rstar \nu'(\rstar)}{2}\right)\varkappa[\Lambda_0'],\label{match:h2p}
\end{align}
\end{widetext}
and
\begin{equation}
\xideform_{1}=\xideform_{\perp}=\mathcal{Q}_{1}=\mathcal{Q}_{\perp}=0.\label{equatorial_deform}
\end{equation}
For the above we have used repeatedly the  identity $[ab]=a^+[b]+b^-[a]$
to find that, e.g., $Q_1[\lambda'^2]=0$.
Observe that because of \eqref{eq:Q1_match}, if $\Lambda_0(\rstar)\neq 0$ then $Q_1=0$,
in all the above relations either $Q_1$ or $\xideform$ appear, never both.

We next compute the differences (the interior and exterior quantities on $r=\rstar$)
of the second order field equations (the set \eqref{eq:dm0}-\eqref{eq:dh2})
to find relations between the jumps of functions \emph{implied by  the field equations} that will have to be
used in combination with
the set of matching conditions above.
Note that the Eulerian equations \eqref{eq:gammas}-\eqref{eq:euler_l2}
only provide information on $r=\rstar$ at the interior
related to $\press_{\ell n}$ and $\press_{\ell p}$, that do not
appear in the matching conditions above.
Likewise, the difference of \eqref{eq:dm0} involves $[v_0']$
and thus does not provide any information that enters the matching conditions above.
We only have to focus on \eqref{eq:dh0}-\eqref{eq:dh2}.

It is convenient to compute first  $[f_\omega]$, which
using the identity $[ab]=a^+[b]+b^-[a]$ yields
\begin{align}
   [f_\omega]=&\varkappa\frac{\rstar^{4}}{3}e^{\lambda(\rstar)}
    \big\{
    -\chi_0(\rstar)p_0(\rstar)(L^2_n-L^2_p)(\rstar)\label{eq:dif_fw}\\
    &-\Lambda_0(\rstar)L_n^2(\rstar)-n_0(\rstar) p_0(\rstar) \A_0(\rstar)(\Omega_n-\Omega_p)^2\big\}.\nonumber
\end{align}
The difference of equation \eqref{eq:dh0} reads
\begin{align}
  &[h_0']-\left(\nu'(\rstar)+\frac{1}{\rstar}\right)[v_0]=\frac{\varkappa}{2} \rstar
  e^{\lambda(\rstar)}\press_0(\rstar)\nonumber\\
    &-\frac{\rstar^3}{6}e^{2\lambda(\rstar)}\varkappa n_0(\rstar)p_0(\rstar)\A_0(\rstar)\left(\Omega_n-\Omega_p\right)^2,\label{eq:dif_h0p}
\end{align}
while the differences of \eqref{eq:v2h2}-\eqref{eq:dh2} read, respectively,
\begin{align}
  [v_2]=&[f_\omega]-[h_2],\label{eq:dif_v2}\\
  [h'_2]+[k_2']=&-\nu'(\rstar)[h_2]+[f_\omega]\left(\frac{1}{\rstar}+\frac{\nu^{\prime}(\rstar)}{2}\right),\label{eq:dif_k2p}
\end{align}
and
\begin{widetext}
\begin{align}
  [h_2']=&
  -\left(\nu^{\prime}(\rstar)+\frac{1}{\rstar \nu^{\prime}(\rstar)}\left(\frac{2}{\rstar}(e^{\lambda(\rstar)}-1)
    -\lambda_-' (\rstar)-\nu'(\rstar)\right)\right)
    [h_2]
    -h^+_2(\rstar)\frac{1}{\nu'(\rstar)}e^{\lambda(\rstar)}\varkappa\Lambda_0(\rstar)\nonumber\\
        &-4\frac{1}{\rstar \nu'(\rstar)}e^{\lambda(\rstar)}([h_2]+[k_2])
          +\frac{1}{2\rstar^2}\left(\frac{2}{\nu^{\prime}(\rstar)}e^{\lambda(\rstar)}
          +\rstar^2\nu^{\prime}(\rstar)\right)[f_\omega],\label{eq:dif_h2p}
\end{align}
\end{widetext}
where we have used the background field equations and the difference
of products in the last equation.

Now we combine the (geometrical) matching conditions with the jumps
of the functions that provide the field equations and find the necessary
and sufficient set of conditions for the matching to exist plus
a relation to obtain the deformation.
We start with the $\ell=0$ sector.
A simple calculation shows that the set of four equations in
\{\eqref{match:h0v0}-\eqref{match:h0p},\eqref{eq:dif_h0p}\}
is equivalent to the set formed by \eqref{eq:dif_h0p} (which is implied by the equations)
plus the two conditions
\begin{align}
  &[h_0]=0,\label{eq:matching_h0}\\
  &[v_0]=\varkappa\frac{\rstar }{\nu^{\prime}(\rstar)}e^{\lambda(\rstar)}  \label{eq:matching_v0}\\
      &\times\bigg\{\frac{1}{3}\rstar^2 e^{\lambda(\rstar)}n_0(\rstar) p_0(\rstar)\A_0(\rstar)(\Omega_n-\Omega_p)^2
      -\press_0(\rstar)
      \bigg\},\nonumber
\end{align}
and the relation
\begin{align}
  &\xideform_{0}\Lambda_0(\rstar)-e^{-\lambda(\rstar)/2}\mathcal{Q}_{0}\Lambda_0'(\rstar)=
    \frac{4}{\nu'(\rstar)}e^{\lambda(\rstar)/2}\nonumber\\
  &\times\bigg\{
  \press_0(\rstar)-\frac{1}{3}\rstar^2 e^{\lambda(\rstar)}n_0(\rstar) p_0(\rstar)\A_0(\rstar)(\Omega_n-\Omega_p)^2\bigg\}\label{eq:deformations_0}
\end{align}
for the deformation (for $\ell=0$).

As for the $\ell=2$ sector, the set of (seven) equations in
\{\eqref{match:k2h2v2},\eqref{match:h2p},\eqref{eq:dif_v2},\eqref{eq:dif_k2p},\eqref{eq:dif_h2p}\} are not all independent and, after some algebra, they can be found to be equivalent to
the set formed by the set of relations \{\eqref{eq:dif_v2},\eqref{eq:dif_k2p},\eqref{eq:dif_h2p}\},
which we recall are implied by the field equations, plus the two conditions
\begin{equation}
  [k_2]=0,\quad [h_2]=0, \label{eq:matching_2}
\end{equation}
and the relation
\begin{align}
  &\xideform_{2}\Lambda_0(\rstar)-e^{-\lambda(\rstar)/2}\mathcal{Q}_{2}\Lambda_0'(\rstar)\nonumber\\
    &=\frac{4}{\nu'(\rstar)}e^{\lambda(\rstar)/2}\left\{\Lambda_0(\rstar)h_2(\rstar)-\frac{1}{\varkappa \rstar^2}[f_\omega]\right\}.\label{eq:deformations_2}
\end{align}

To sum up, the necessary and sufficient conditions
for the matching of the interior and exterior problems
at second order are \eqref{eq:matching_h0} and  \eqref{eq:matching_v0}
together with \eqref{eq:matching_2}. Moreover, the deformation quantities
are given by \eqref{eq:deformations_0} and \eqref{eq:deformations_2}.
It is crucial that  \eqref{eq:matching_v0} exhibits a jump which
has been overlooked in all the previous literature on two-fluid models,
and is, of course, directly related to the correction to the
Hartle-Thorne model presented in \cite{ReinaVera2015} (see also \cite{MRV2}).
That jump has consequences on the calculation of $\delta M$,
and the expression given in \cite{Andersson_Comer_2001}
needs to be corrected. We will provide that expression later,
when we compute the global solution of the second order global
problem. 

\subsection{Solving the second order global problem}

We separate the second order problem into the $\ell=0$ and $\ell=2$ sectors.

\subsubsection{$\ell=0$}
In order to integrate the system in the interior region we
work with
the set of functions $\{v^+_0,\press_{0n},\press_{0p}\}$, as follows.
We differentiate the two equations in \eqref{eq:gammas} and \eqref{eq:euler_l2}
and substitute $h_0^\prime$ from \eqref{eq:dh0} to obtain (we drop the $+$ indicators)
\begin{widetext}
\begin{align}
  \press_{0n}^{\prime}=-\frac{\nu'}{2}\press_{0n}+\mu_0
  &\bigg\{-\frac{\varkappa}{2} re^\lambda
    (n_0\press_{0n}+p_0 \press_{0p})
    +\frac{\varkappa}{6}r^3e^{\lambda-\nu}n_0 p_0\A_0(\Omega_n-\Omega_p)^2\nonumber\\
    &-\left(\nu^{\prime}+\frac{1}{r}\right)v_0
    +\frac{1}{3}\left(r^2e^{-\nu} L_n^2 \right)^{\prime}+\frac{1}{12}r^3e^{-\nu}{\omega^{\prime}}^2\bigg\},\label{eq:dP0n}\\
  \press_{0p}^{\prime}=-\frac{\nu'}{2}\press_{0p}+\chi_0
  &\bigg\{-\frac{\varkappa}{2} re^\lambda
    (n_0\press_{0n}+p_0 \press_{0p})
    +\frac{\varkappa}{6}r^3e^{\lambda-\nu}n_0 p_0\A_0(\Omega_n-\Omega_p)^2\nonumber\\
  &-\left(\nu^{\prime}+\frac{1}{r}\right)v_0
    +\frac{1}{3}\left(r^2e^{-\nu} L_p^2 \right)^{\prime}+\frac{1}{12}r^3e^{-\nu}{\omega^{\prime}}^2\bigg\}.\label{eq:dP0p}
\end{align}
\end{widetext}

The system to integrate is thus given by Eqs.~\eqref{eq:dm0},~\eqref{eq:dP0n},~\eqref{eq:dP0p}.
The conditions we impose at the origin are $\press_{0n}(0)=0$ and $\press_{0p}(0)=0$,
which correspond to the conditions $\eta_0(0)=\Phi_0(0)=0$
in \cite{Andersson_Comer_2001} (c.f.~Eq.~\eqref{eq:def_pressl}).
{Operationally we set $\Omega_p=1$ and $\Omega_n=\Delta$ in the equations. The solutions thus provide, in fact, the set
$\{\hat v^+_0,\hat \press_{0n},\hat \press_{0p}\}:=\{v^+_0/\Omega_p^2,\press_{0n}/\Omega_p^2,\press_{0p}/\Omega_p^2\}$ in terms of $\Delta$.
}
The equations, given the known behaviour of $\lambda$ and $\nu$ near the
origin as shown in Section \ref{sec:background_comp},
present a structure of the poles at the origin
that imply that the only solution to the homogenous problem is the trivial one.
This means that if there is a bounded solution, that is
the unique solution $\{\hat v^+_0(r),\hat \press_{0n}(r),\hat \press_{0p}(r)\}$.
Moreover, $\hat v^+_0$ is $O(r^4)$ and $\hat \press_{0n}$ and $\hat \press_{0p}$ are $O(r^2)$.
This is analogous to the perfect fluid case (see \cite{MRV2}).

It is important to note
that the combination $\chi_0\times$\eqref{eq:dP0n}-$\mu_0\times$\eqref{eq:dP0p} can be readily integrated to get
\begin{equation}
  \label{eq:diff_P}
  \chi_0\hat \press_{0 n}-\mu_0  \hat \press_{0 p}=\mu_0\chi_0\frac{r^2}{3e^\nu}
  \left(2\hat\omega_+(1-\Delta)+\Delta^2-1\right),
\end{equation}
after using that $\hat\press_{0n}$ and $\hat\press_{0p}$ vanish at the origin.
Introducing both equations from \eqref{eq:gammas} in this relation
it is straightforward to obtain
\[
  \gamma_n=\gamma_p.
  \]
We will use these relations later.

Once the solution $\{\hat v^+_0(r),\hat \press_{0n}(r),\hat \press_{0p}(r)\}$
is found, we use the matching conditions \eqref{eq:matching_h0}-~\eqref{eq:matching_v0} (conveniently divided by $\Omega_p^2$)
to obtain the value of $\delta M$,
which is the only constant on the exterior solution \eqref{eq:v0_h0_ext}
for $\ell=0$ that needs to be determined.
We thus obtain
\begin{widetext}
\begin{align}
  \label{eq:deltaM}
  \delta\hat{M}:=&\frac{1}{\Omega_p^2}\delta M=\frac{1}{\Omega_p^2}\bigg(\frac{J^2}{\rstar^3}+(\rstar-2M)\left(v^+_0(\rstar)-[v_0]\right)\bigg)\nonumber\\
  =&\frac{\hat{J}^2}{\rstar^3}+(\rstar-2M)\hat v^+_0(\rstar)-\varkappa\frac{\rstar(\rstar-2M)}{\nu^{\prime}(\rstar)}e^{\lambda(\rstar)}
      \bigg\{\frac{1}{3}\rstar^2 e^{\lambda(\rstar)}n_0(\rstar) p_0(\rstar)\A_0(\rstar)(\Delta-1)^2
      -\hat\press_0(\rstar)
      \bigg\}.
\end{align}
\end{widetext}
Observe that expression (60) in \cite{Andersson_Comer_2001} needs
to be corrected with the term containing the factor within curly brackets.

With the value of $\delta \hat{M}$ we have thus determined $\hat h^-_0(r){:=h^-_0(r)/\Omega^2_p}$
and therefore also $\hat h^+_0(\rstar)=\hat h_0^-(\rstar)$ because
of $[h_0]=0$.
It only remains to evaluate either one in \eqref{eq:gammas}
on $r=\rstar$ using the values of the integrated
functions $\hat\press_{0p}(\rstar)$ (or $\hat\press_{0n}(\rstar)$)
to obtain the value of $\hat\gamma_p=\hat\gamma_n$.
The whole function $h_0^+(r)$ is then obtained by isolating it from
either relation in \eqref{eq:gammas}.

\subsubsection{The total mass}
The computation of the total mass
(using e.g. the approach in \cite{Laarakkers_1999}) of the family of geometries given by $g_\pert$
with \eqref{g_stax} and \eqref{NKV} at $r\to\infty$, that depends on the background configuration
and the central values of $n_0$ and $p_0$, leads to
\begin{align}
  M_T(n_0,p_0)=&M(n_0,p_0)+\pert^2 \delta M(n_0,p_0)\nonumber\\
  =&M(n_0,p_0)+\tOmega_p^2 \delta \hat{M}(n_0,p_0),\label{M_T}
\end{align}
using $\pert=\tOmega_p/\Omega_p$, with
$\delta \hat{M}$ given by \eqref{eq:deltaM}.

\subsubsection{$\ell=2$}
In the $\ell=2$ sector the problem is set for the pair of functions $\{\hat k_2,\hat h_2\}$,
and the system to integrate in the interior region is given by the set
of equations \{\eqref{eq:dk2dh2},\eqref{eq:dh2}\} (setting
$\Omega_p=1$ and $\Omega_n=\Delta$ ). 
This time, the structure of the poles at $r=0$ implies that the
bounded solutions to the homogeneous problem are all proportional
to some homogeneous solution, that we shall denote by $\{k_2{}_H, h_2{}_H\}$.
Therefore, the general solution of the interior problem is given by
\[
  \hat k_2(r)=A k_2{}_H(r)+k_{2P}(r),\quad \hat h_2(r)=A h_2{}_H(r)+h_{2P}(r),
\]
with $A\in\mathbb{R}$, where $k_{2P}(r)+h_{2P}(r)$ and  $h_{2P}(r)$ are particular solutions
of \{\eqref{eq:dk2dh2},\eqref{eq:dh2}\} (with $\Omega_p=1$ and
$\Omega_n=\Delta$).
Recalling that the exterior solution is given by the expressions \eqref{eq:h2_ext}
and \eqref{eq:k2_ext}, we have two constants to fix, namely $A$ and $\hat \constK:=\constK/\Omega_p^2$.
These two constants are determined by the (only) two matching conditions \eqref{eq:matching_2}
in the $\ell=2$ sector, explicitly
\begin{align*}
  A k_2{}_H(\rstar)+k_{2P}(\rstar)=\hat k_2^-(\rstar),\\
  A h_2{}_H(\rstar)+h_{2P}(\rstar)=\hat h_2^-(\rstar),
\end{align*}
with $k_2^-$ and $h_2^-$ given by  \eqref{eq:h2_ext} and \eqref{eq:k2_ext}.
Once we have determined the pair $\{\hat k_2(r),\hat h_2(r)\}$, the perturbation metric
function $\hat v_2(r)$ is determined by the algebraic equation \eqref{eq:v2h2}.

\subsubsection{Quadrupole moment}

Having obtained the value of the constant $\hat \constK$
in the exterior solution, {we can compute
the quadrupole moment of the star using the procedure
in \cite{Laarakkers_1999} (see also~\cite{Hartle3}; observe this is $-Q$ as defined in \cite{Andersson_Comer_2001}) to obtain
\[
  Q^S=\pert^2\left(\frac{8\constK M^3}{5}+\frac{J^2}{M}\right)=\tOmega_p^2\left(\frac{8\hat \constK M^3}{5}+\frac{\hat{J}^2}{M}\right).
\]}
\section{Deformation}
\label{sec:deformations}

Once we have the whole perturbed solution (at second order) in terms of the
perturbation functions $\{v,h,k\}$,
it only remains to determine the deformation
of the surface of the star with respect to the spherical hypersurface at $r=\rstar$.

The obtaining of the deformation in the perturbation scheme we have developed in
the previous sections amounts to the computation of the function
$\xideform$ (and $Q_1$ at first order) by means
of \eqref{equatorial_deform}, \eqref{eq:deformations_0} and \eqref{eq:deformations_2}
together with \eqref{eq:Q1_match_}.
Therefore, the procedure only determines the deformation,
by means of $\xideform_{0}$  and $\xideform_{2}$,
if $\Lambda_0(\rstar)\neq 0$.
This is analogous to what happens in the perfect fluid case with barotropic equation
of state, as shown in \cite{ReinaVera2015} (see also \cite{MRV2}).
If $\Lambda_0(\rstar)\neq 0$, then $Q_1=0$ by \eqref{eq:Q1_match_} and $\xideform$
reads
\[
  \xideform(\theta)=\xideform_{0}+\xideform_{2}P_2(\cos\theta)
\]
with
\begin{align}
  &\xideform_{0}
    =\frac{4e^{\lambda(\rstar)/2}}{\nu'(\rstar)\Lambda_0(\rstar)}\label{eq:deformations_0_final}\\
    &\times\bigg\{
    \press_0(\rstar)-\frac{1}{3}\rstar^2 e^{\lambda(\rstar)}n_0(\rstar) p_0(\rstar)\A_0(\rstar)(\Omega_n-\Omega_p)^2\bigg\},\nonumber\\
    &\xideform_{2}=\frac{4e^{\lambda(\rstar)/2}}{\nu'(\rstar)}\bigg\{h_2(\rstar)-\frac{1}{\varkappa \rstar^2\Lambda_0(\rstar)}[f_\omega]\bigg\},\label{eq:deformations_2_final}
\end{align}
which, observe, are constants.
Using equations \eqref{eq:euler_l2} and \eqref{eq:dif_fw}  it is straightforward to
obtain the equality
\begin{align*}
  &n_0(\rstar)\press_{2n}(\rstar)+p_0(\rstar)\press_{2p}(\rstar)=h_2\Lambda_0(\rstar)
  -\frac{[f_\omega]}{\varkappa\rstar^2}\\
  &-\frac{1}{3}\rstar^2e^{\lambda(\rstar)}n_0(\rstar)p_0(\rstar)\A_0(\rstar)(\Omega_n-\Omega_p)^2
\end{align*}
that allows us to conveniently reexpress the above pair of equations in the compact form
\begin{align}
  &\xideform_{\ell}
    =\frac{4e^{\lambda(\rstar)/2}}{\nu'(\rstar)\Lambda_0(\rstar)}\bigg\{
    \press_\ell(\rstar)
    \nonumber\\
    &-\frac{1-\ell}{3}\rstar^2 e^{\lambda(\rstar)}n_0(\rstar) p_0(\rstar)\A_0(\rstar)(\Omega_n-\Omega_p)^2\bigg\}.\label{eq:deformations_l_final}
\end{align}

Let us stress that in the perfect fluid case with barotropic equation
of state the factor $E(\rstar)$ (the energy density at the boundary)
appears multiplying both left and right hand sides
of the equations analogous to \eqref{eq:deformations_0} and \eqref{eq:deformations_2}
and thus disappears from the denominators in the expressions for the deformation.
In the present case
the fact that we need $\Lambda_0(\rstar)\neq 0$ is made explicit in the expressions.
Despite that, as argued in \cite{ReinaVera2015},
the perturbative procedure eventually determines the deformation if it is continued
further to higher orders, providing the same result.
Moreover,  
it is shown that the result of that procedure is equivalent to use
the argument presented in \cite{Hartle1967}, based on the Newtonian approach
in \cite{Chandra_Poly_Newton_1933} (see \cite{ReinaVeranote}),
in which the perturbative configuration
is assumed to be the second order expansion of a given exact configuration. Observe
that following a strict perturbative procedure one cannot ensure the sum of
all the orders, and therefore the existence of the full configuration.

The argument to obtain the deformation using the full configuration
follows the exact result, see Section \ref{sec:surface_star},
by which the surface of the star is determined by
$\Psi(r,\theta)=0$, a curve on the plane $\{r,\theta\}$.
Observe that in the perfect fluid case
(no equation of state needed) that corresponds to $P(r,\theta)=0$
(where $P$ is the pressure).
The computation in \cite{Hartle1967}, however, is made in terms of
the surface levels of the energy density function $E$.
Nevertheless, since it is assumed a barotropic equation of state of the form $E(P)$,
those surface levels correspond also to the surface levels of the pressure
function $P$ and the result is therefore the same.

In the present case, however, the computation
in \cite{Andersson_Comer_2001} (see Eq (33) to (39) there) is made using the function $\Lambda$,
while the correct computation requires using the function $\Psi$.
Following the argument in \cite{Andersson_Comer_2001} but replacing
$\Lambda$ by $\Psi$, the deformation to second order $\xi(\rstar,\theta)$
is defined by the second order term of the total deformation (we add the parameter $\pert$ here)
\[
\tilde r_\pert(r,\theta)=r +\pert^2\xi(r,\theta) + O(\pert^3),
\]
determined
by $\Psi_\pert(\tilde r_\pert(r,\theta),\theta)=\Psi_0(r)$ and thus given by
\begin{equation}
  \label{eq:AC_xi}
  \Psi_\pert(r,\theta)=\Psi_0(r)-\Psi_0'(r)\pert^2\xi(r,\theta)+O(\pert^3)
\end{equation}
(mind, for all $r\in(0,\rstar)$).
On the other hand,
the expasion of $\Psi_\pert(r,\theta)=\Psi(n_\pert(r,\theta)^2,p_\pert(r,\theta)^2,x_\pert(r,\theta)^2)$ in $\pert$ is given by (c.f. equation (106) in \cite{Andersson_Comer_2001})
\begin{widetext}
\begin{align}
  \Psi_\pert=&\Psi_0+\pert^2\left\{\frac{}{}(n_0 \B^0_0+p_0\A^0_0)n_0\eta+(p_0 \C^0_0+n_0\A^0_0)p_0\Phi
  \right.\nonumber\\
             &\left.+\frac{1}{2}n_0p_0\left(\A_0+2n^2_0\frac{\partial \A_0}{\partial n_0^2}
               +2p^2_0\frac{\partial \A_0}{\partial p_0^2}+2n_0p_0\frac{\partial \A_0}{\partial x_0^2}\right)r^2\sin^2\theta e^{-\nu}(\Omega_n-\Omega_p)^2\right\}.
\end{align}
Introducing this into \eqref{eq:AC_xi},
using $\sin^2\theta=2/3(P_0(\cos\theta)-P_2(\cos\theta))$,
together with  \eqref{eq:psi'} and \eqref{def:DnDp} we find
\[
  \xi(r,\theta)=\xi_0(r)+\xi_2(r)P_2(\cos\theta)
\]
with
\begin{align}
  \xi_\ell=&\frac{2}{\nu'}\frac{1}{\Psi_0-\Lambda_0}
  \left\{(n_0 \B^0_0+p_0\A^0_0)n_0\eta_\ell+(p_0 \C^0_0+n_0\A^0_0)p_0\Phi_\ell
  +(1-\ell)n_0p_0 (\mathcal{D}_n+\mathcal{D}_p-\A_0)\frac{r^2}{3 e^\nu}(\Omega_n-\Omega_p)^2\right\}\nonumber\\
  =&\frac{2}{\nu'}\frac{1}{\Psi_0-\Lambda_0}
  \left\{(n_0 p_0)\mathfrak{A}\vecbl+(1-\ell)n_0p_0 (\mathcal{D}_n+\mathcal{D}_p-\A_0)\frac{r^2}{3 e^\nu}(\Omega_n-\Omega_p)^2\right\}.\label{eq:xis_etaphi}
\end{align}
  \end{widetext}
Using \eqref{eq:def_pressl}, in terms of $\press_{\ell n}$ and
$\press_{\ell p}$, this reads
\begin{align}
  &\xi_\ell(r)=\frac{2}{\nu'(r)}\frac{1}{\Psi_0(r)-\Lambda_0(r)}
  \bigg\{n_0(r)\press_{\ell n}(r)+p_0(r)\press_{\ell p}(r)\nonumber\\
  &-(1-\ell)n_0(r) p_0(r) \A_0(r)\frac{r^2}{3 e^{\nu(r)}}(\Omega_n-\Omega_p)^2\bigg\}.\label{eq:xis_r}
\end{align}
Finally, evaluating $\xi_\ell(r)$ on $r=\rstar$, using that $\Psi_0(\rstar)=0$
and $\nu(\rstar)=-\lambda(\rstar)$,
plus the auxiliary definition \eqref{def:Pl},
we directly obtain
\begin{align}\label{eq:xis_S}
  &\xi_\ell(\rstar)=-\frac{2}{\nu'(\rstar)\Lambda_0(\rstar)}
  \bigg\{
  \press_\ell(\rstar)\nonumber\\
    &-(1-\ell)n_0(\rstar) p_0(\rstar) \A_0(\rstar)\frac{\rstar^2}{3}e^{\lambda(\rstar)}(\Omega_n-\Omega_p)^2\bigg\},
  \end{align}
and therefore, by comparing with \eqref{eq:deformations_l_final},
\[
  \xi_\ell(\rstar)=-\frac{1}{2}e^{-\lambda(\rstar)/2}\xideform_{\ell}.
\]
{The factors in this expression account for the $1/2$ that ought to be included in $\xi$ for being a second order quantity (the $\frac{1}{2}\pert^2$ factor),
while $-e^{-\lambda/2}$ corresponds to the direction 
and normalisation of the normal to the (background) surface $\normal_0=-e^{-\lambda/2}\partial_r$ (see Subsection \ref{pert_scheme}).}

Let us stress that the expressions of $\xi_\ell(\rstar)$ contain also $\Lambda_0(\rstar)$
in the denominators. The advantage of this procedure, assuming that the
exact configuration exists, is that \eqref{eq:xis_r} hold for all $r\in(0,\rstar)$,
and therefore $\xi_\ell(\rstar)$ are to be obtained as
the limits of $\xi_\ell(r\to\rstar)$ \emph{if they exist}. It is interesting to note that if $\A_0=0$ then $\quotomega$ does not affect the deformation.

As mentioned, the procedure to obtain the deformation
presented in \cite{Andersson_Comer_2001}
is just the analogue to the above but using $\Lambda$ instead of $\Psi$.
Explicitly, the deformation is claimed in  \cite{Andersson_Comer_2001}
to be the function  $\xiAC(r,\theta)$ that solves the equation (we add the perturbation parameter $\pert$)
\begin{equation}
  \Lambda_\pert(r,\theta)=\Lambda_0(r)-\Lambda_0^{\prime}\pert^2\xiAC(r,\theta)+ O(\pert^3).\label{eq:def_AC}
\end{equation}
This function $\xiAC$ here is denoted by $\xi$ in \cite{Andersson_Comer_2001}.
Next we show that
the outcome is just different, and therefore $\xiAC$ does not provide
the deformation.
From \eqref{eq:def_AC} and the expansion of
$\Lambda_\pert(r,\theta)=\Lambda(n_\pert(r,\theta)^2,p_\pert(r,\theta)^2,x_\pert(r,\theta)^2)$ in $\pert$,
which is given by (c.f. equation (105) in \cite{Andersson_Comer_2001})
\begin{align}
     \Lambda_\pert =\Lambda_0 -&\pert^2\bigg\{\mu_0 n_0 \eta + \chi_0 p_0 \Phi \nonumber\\
     &+ \frac{1}{2 e^\nu}r^2\sin^2\theta\mathcal{A}_0 n_0 p_0(\Omega_n - \Omega_p)^2\bigg\},
\end{align}
and using  \eqref{eq:Lambda'} for $\Lambda_0'$ and \eqref{eq:euler_back1}-\eqref{eq:euler_back2}, we obtain
\[
  \xiAC(r,\theta)=\xiAC_0(r)+\xiAC_2(r)P_2(\cos\theta)
\]
with
\begin{widetext}
\begin{align}
  \xiAC_\ell(r) =&\frac{2}{\nu'(r)}\frac{e^{\nu(r)/2}\det\mathfrak{A}(r)}{\C^0_0(r)\mu_\infty^2-2\mu_\infty\chi_\infty\A^0_0(r)+\B^0_0(r)\chi_\infty^2}
  \nonumber\\
  &\times
  \bigg\{ \mu_\infty(r) n_0(r) \eta_\ell(r) + \chi_\infty(r) p_0(r) \Phi_\ell(r)
  +(1-\ell) \frac{r^2}{3 e^{\nu(r)}}\mathcal{A}_0(r) n_0(r) p_0(r)(\Omega_n - \Omega_p)^2\bigg\}.
\end{align}
In terms of $\press_{\ell n}$ and $\press_{\ell p}$ using \eqref{eq:def_pressl},
this reads (we avoid to make the $r$ dependence explicit now)
\begin{align}
  \xiAC_\ell =&\frac{2}{\nu'}\frac{e^{\nu/2}}{\C^0_0\mu_\infty^2-2\mu_\infty\chi_\infty\A^0_0+\B^0_0\chi_\infty^2}\biggl\{(\mu_\infty\C_0^0-\chi_\infty\A_0^0)\press_{\ell n}
                +(\chi_\infty\B_0^0-\mu_\infty\A_0^0)\press_{\ell p}\nonumber\\
              &+(1-\ell) \frac{r^2}{3 e^{\nu}}(\Omega_n - \Omega_p)^2
                \left(\det\mathfrak{A}\mathcal{A}_0 n_0 p_0
                -p_0\mathcal{D}_n(\mu_\infty\C_0^0-\chi_\infty\A_0^0)
                -n_0\mathcal{D}_p(\chi_\infty\B_0^0-\mu_\infty\A_0^0)\right)\biggl\}.\label{eq:xi_l_AC}
\end{align}
\end{widetext}
In the particular (numerical) case studied in \cite{Andersson_Comer_2001},
that we will retake in the following Section \ref{numerical},
the function $\Lambda$ is such that $\A_0=0$ (and therefore $\mathcal{D}_n=\mathcal{D}_p=0$
by definition), and it is assumed $\chi_\infty=\mu_\infty$.
For any given model with those two assumptions, equation \eqref{eq:xis_r} using
\eqref{eq:E+P}, and the above expressions
simplify to
\begin{align*}
  &\xi_\ell(r) =\frac{2}{\nu'(r)}\frac{e^{\nu(r)/2}}{\mu_\infty }\frac{\press_\ell(r)}{n_0(r)+p_0(r)},
  \end{align*}
  and
  \begin{align*}
  &\xiAC_\ell(r)=\frac{2}{\nu'(r)}\frac{e^{\nu(r)/2}}{\mu_\infty }\frac{(\C_0^0-\A_0^0)\press_{\ell n}
                  +(\B_0^0-\A_0^0)\press_{\ell p}}{\C^0_0(r)-2\A^0_0(r)+\B^0_0(r)}.
\end{align*}
Therefore, in general $\xi_\ell(r)\neq \xiAC_\ell(r)$
even in that case. To check the equality, or not,
of the limits $\xi_\ell(r\to \rstar)$ and $\xiAC_\ell(r\to\rstar)$ requires
the integration of the whole system, and we leave that to Section \ref{numerical}.
However, if we also demand $\Omega_n=\Omega_p$ then
\eqref{eq:euler_l2} and \eqref{eq:diff_P} with $\mu_\infty=\chi_\infty$ lead to
\[
  \press_{\ell n}(r)-\press_{\ell p}(r)=0,
\]
so that $\press_\ell=(n_0+p_0)\press_{\ell n}$ and therefore 
\[
  \xi_\ell=\xiAC_\ell=\frac{2}{\nu'(r)}\frac{e^{\nu(r)/2}}{\mu_\infty }\press_{\ell n}.
\]
Let us stress that if one imposes
$\mu_\infty=\nu_\infty$ and $\Omega_n=\Omega_p$ then $\xi_\ell=\xiAC_\ell$ irrespective of the form of $\Lambda$.

\section{Kepler limit}
\label{Kepler}

The mass-shedding (Kepler) limit $\tOmega_K$ is the maximum angular velocity of a particle rotating at the equator of the star as seen by an observer at infinity. Therefore, this sets the limit on the angular momentum of the fluids forming the star. For the present two-fluid model, the Kepler limit will correspond to the limit of the fastest-spinning component, i.e. $\tOmega_K=\max(\tOmega_n,\tOmega_p)$.

As stated in \cite{Andersson_Comer_2001} the Kepler limit is given by
\begin{equation}
    \tOmega_K = \frac{Nv}{\sqrt{K}} + N^\phi,\label{eq:OmegaK1}
\end{equation}
where
\begin{equation*}
    v = \frac{K^{3/2}{N^{\phi}}^\prime}{NK^\prime} + \sqrt{\frac{2KN^\prime}{NK^\prime} + \left(\frac{K^{3/2}{N^{\phi}}^\prime}{NK^\prime}\right)^2},\label{eq:v1}
\end{equation*}
in terms of the functions in \eqref{g_stax}.

\begin{figure*}[t!]
        \includegraphics[width=0.48\textwidth]{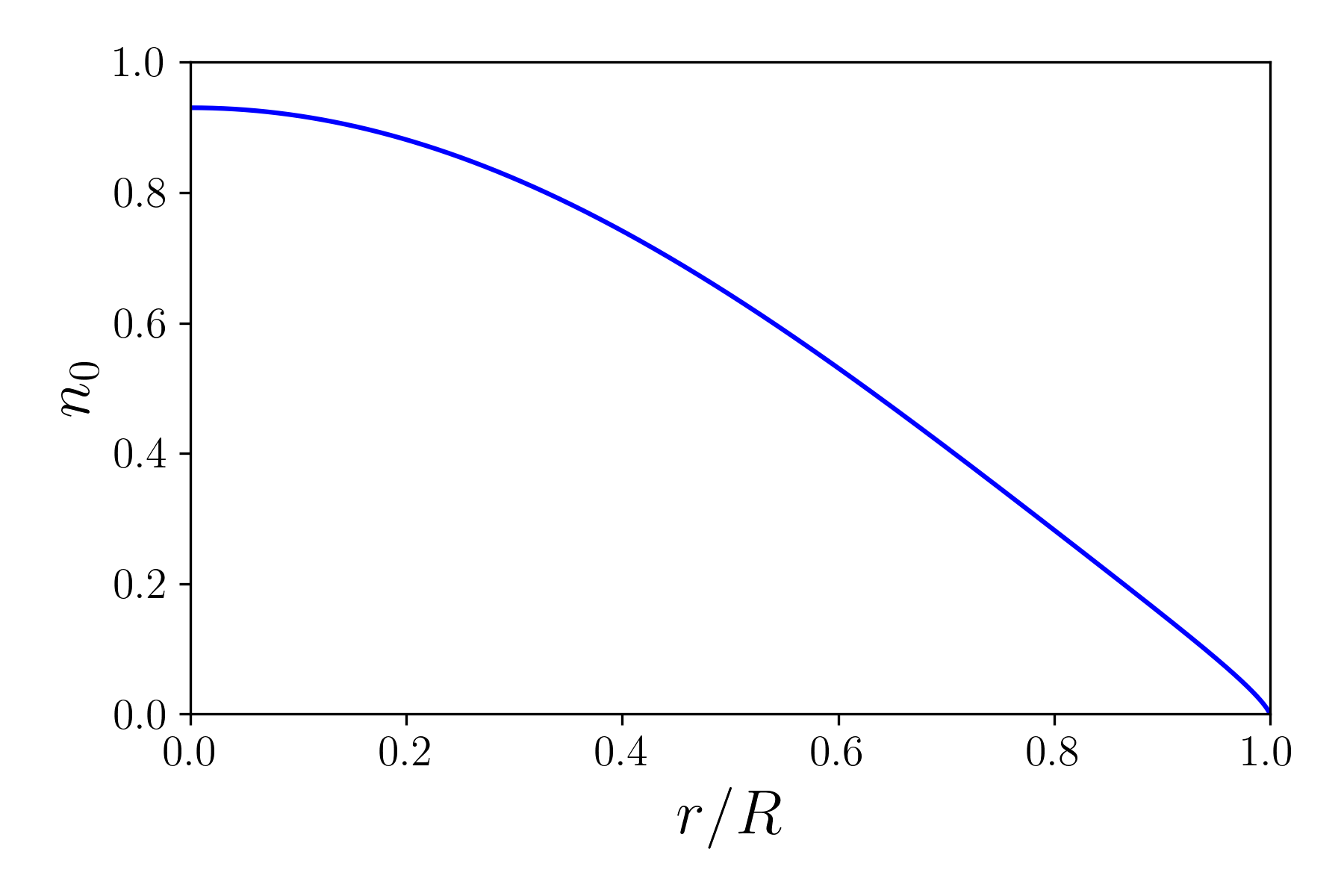}
        \includegraphics[width=0.48\textwidth]{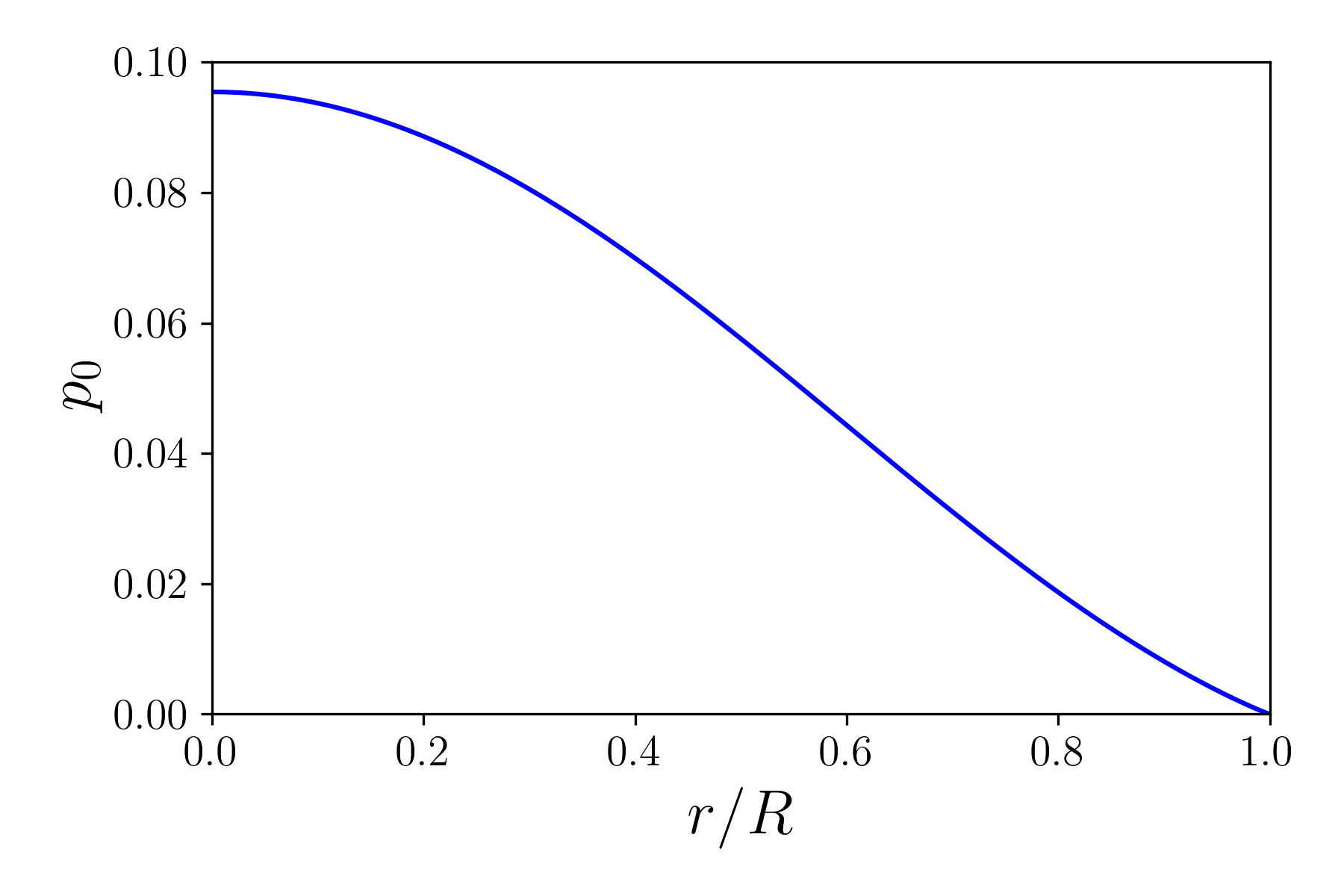}
        \\
        \includegraphics[width=0.48\textwidth]{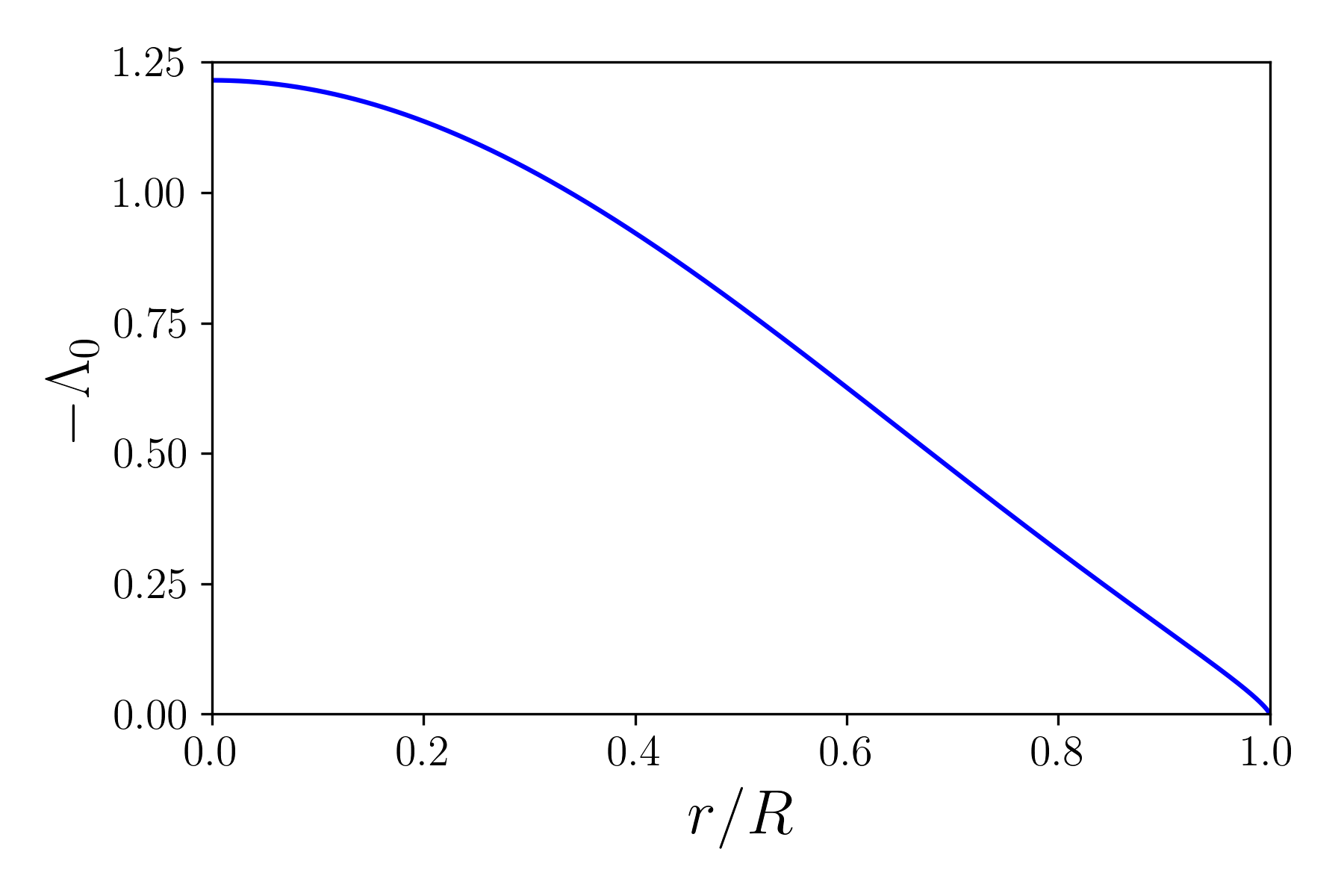}
        \includegraphics[width=0.48\textwidth]{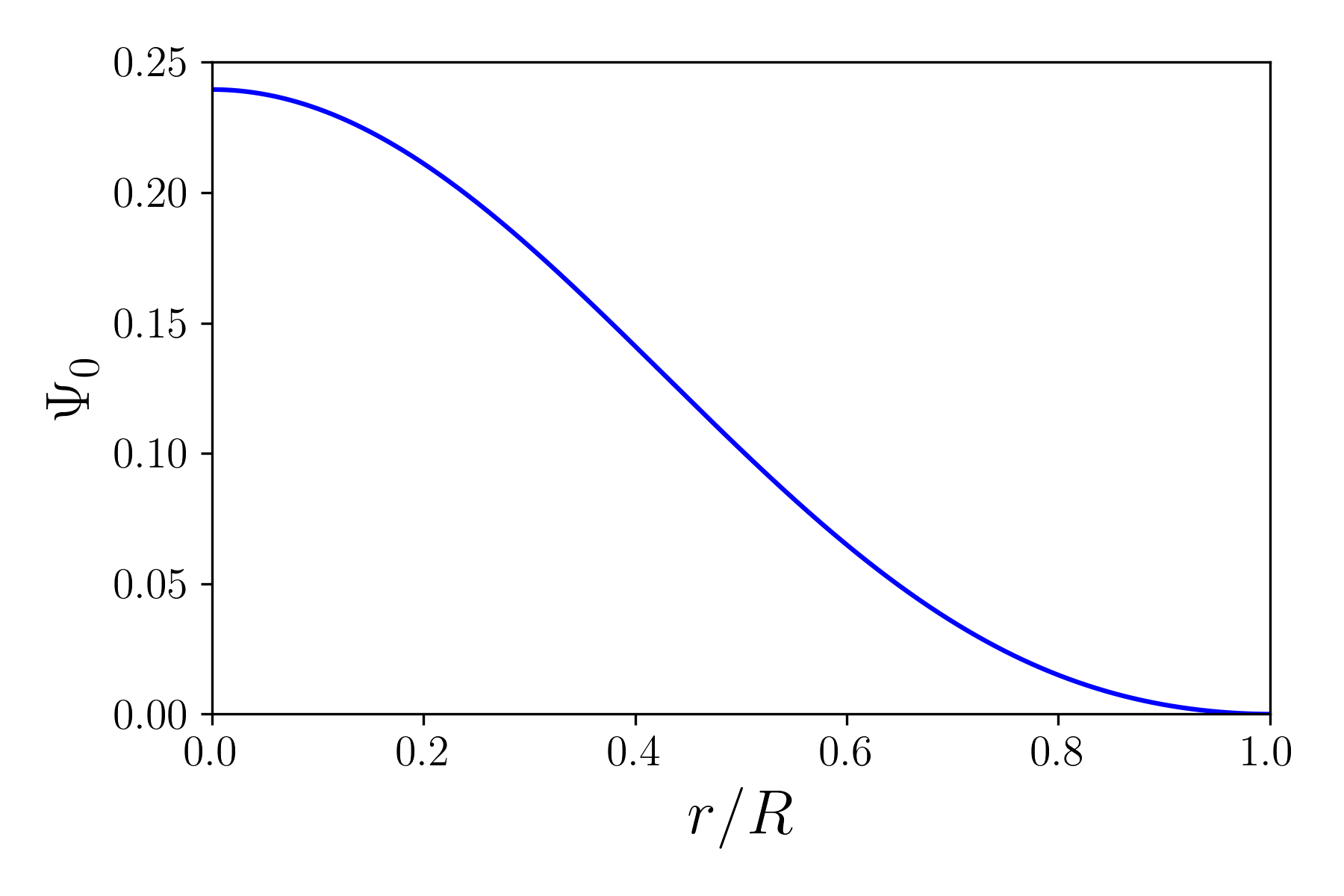}
    \caption{Normalized radial profiles of relevant quantities of the background configuration for the AC model (in code units): number density of neutrons $n_0(r)$ (top-left), number density of protons, $p_0(r)$ (top-right), energy density, $-\Lambda_0(r)$ (bottom-left), and generalized pressure, $\Psi_0(r)$ (bottom right). All four functions vanish at the boundary of the star.} 
    \label{fig:densities}
\end{figure*}
Using \eqref{NKV} this relation reads
\begin{align}
    &\tOmega_K=\frac{e^{\nu/2}}{r}\sqrt{\frac{r\nu^\prime}{2}} + \pert\left(\omega+\frac{r\omega^\prime}{2}\right) + \pert^2e^{\nu/2}\sqrt{\frac{\nu^\prime}{2r}}\nonumber\\
    \times&\left[h-k+\frac{h^\prime}{\nu^\prime}-\frac{rk^\prime}{2}+\frac{r^3(\omega^\prime)^2}{4\nu^\prime e^\nu}\right]+O(\pert^3),\label{eq:OmegaK2}
\end{align}
which, evaluated on the equator ($\theta=\pi/2$) of the perturbed boundary of the star, $r=\rstar+\pert^2\xi(\rstar)$,
yields
\begin{widetext}
\begin{align}
    &\tOmega_K = \sqrt{\frac{M}{\rstar^3}} - \tOmega_p \frac{\hat{J}}{\rstar^3} + \tOmega_p^2\sqrt{\frac{M}{\rstar^3}}\left\{\frac{\delta\hat{M}}{2M} + \frac{(\rstar+3M)(3\rstar-2M)}{4\rstar^4M^2}\hat{J}^2 - \frac{3}{4\rstar}
    (2\hat \xi_0(\rstar)-\hat \xi_2(\rstar))
    +\beta
    \hat{\constK}
    \right\}+O(\pert^3)\label{eq:OmegaK3}
\end{align}
\end{widetext}
after using $\pert=\tOmega_p/\Omega_p$,
and where 
\begin{align*}
    \beta=&\frac{3(\rstar^3-2M^3)}{4M^3}\log\left(1-\frac{2M}{\rstar}\right)\\
    &+\frac{3\rstar^4-3\rstar^3M-2\rstar^2M^2-8\rstar M^3+6M^4}{2\rstar M^2(\rstar-2M)}.
\end{align*}
The expression \eqref{eq:OmegaK3} (see also \cite{Kheto2015}) differs slightly from expression (77) in \cite{Andersson_Comer_2001} in the square root factor at second order,
the sign in front of $\xi_2$, and its denominator with $R^2$, which may simply be typos. Since the Kepler limit will correspond to the fastest rotating fluid, for $\Delta>1$ we have $\tOmega_K=\tOmega_n=\Delta\tOmega_p$, whereas for $\Delta<1$ we have $\tOmega_K=\tOmega_p$. In any case, we solve the quadratic equation for $\tOmega_p$, and take the smallest result among the two solutions.

\begin{figure}[t]
        \includegraphics[width=\linewidth]{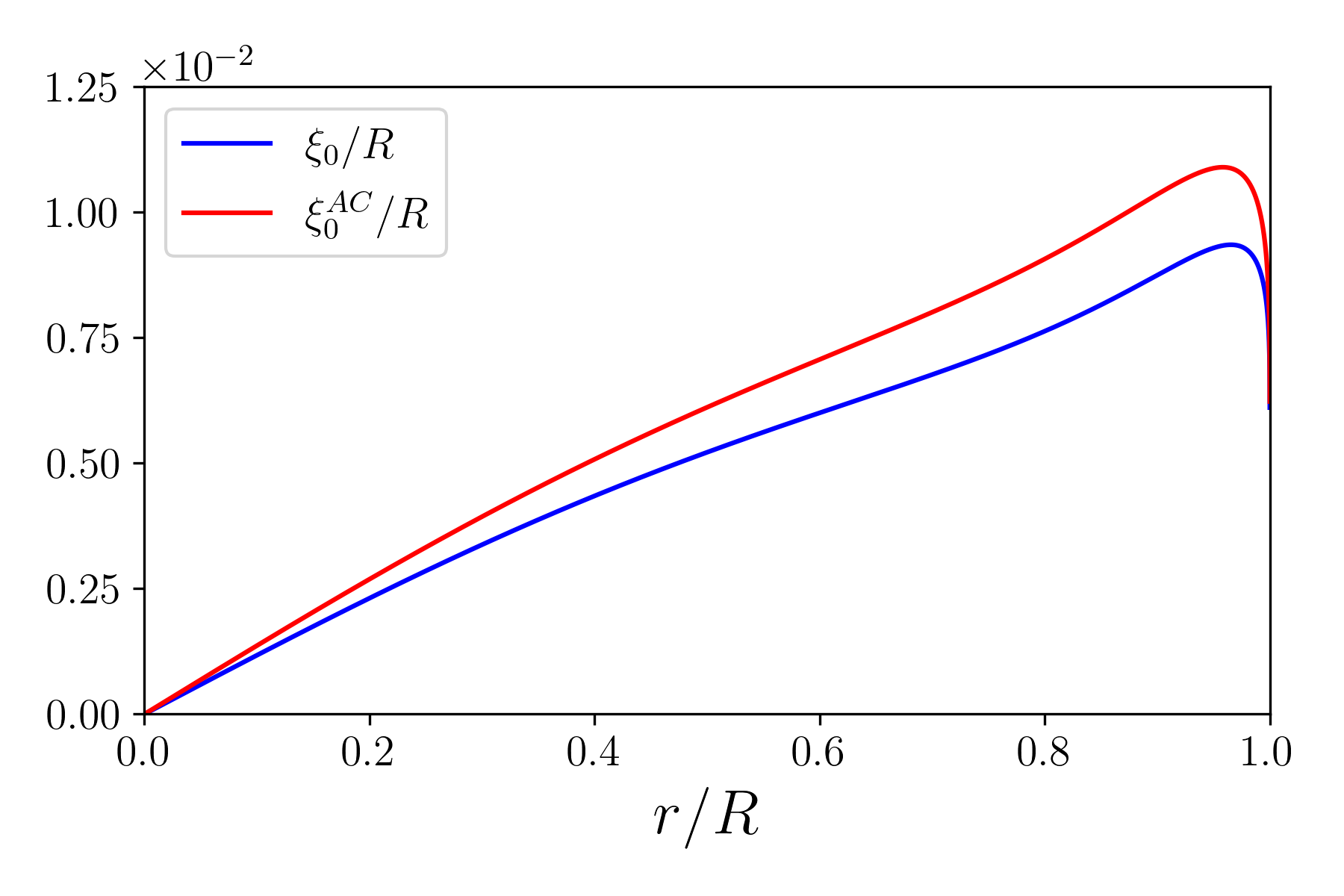}
        \includegraphics[width=\linewidth]{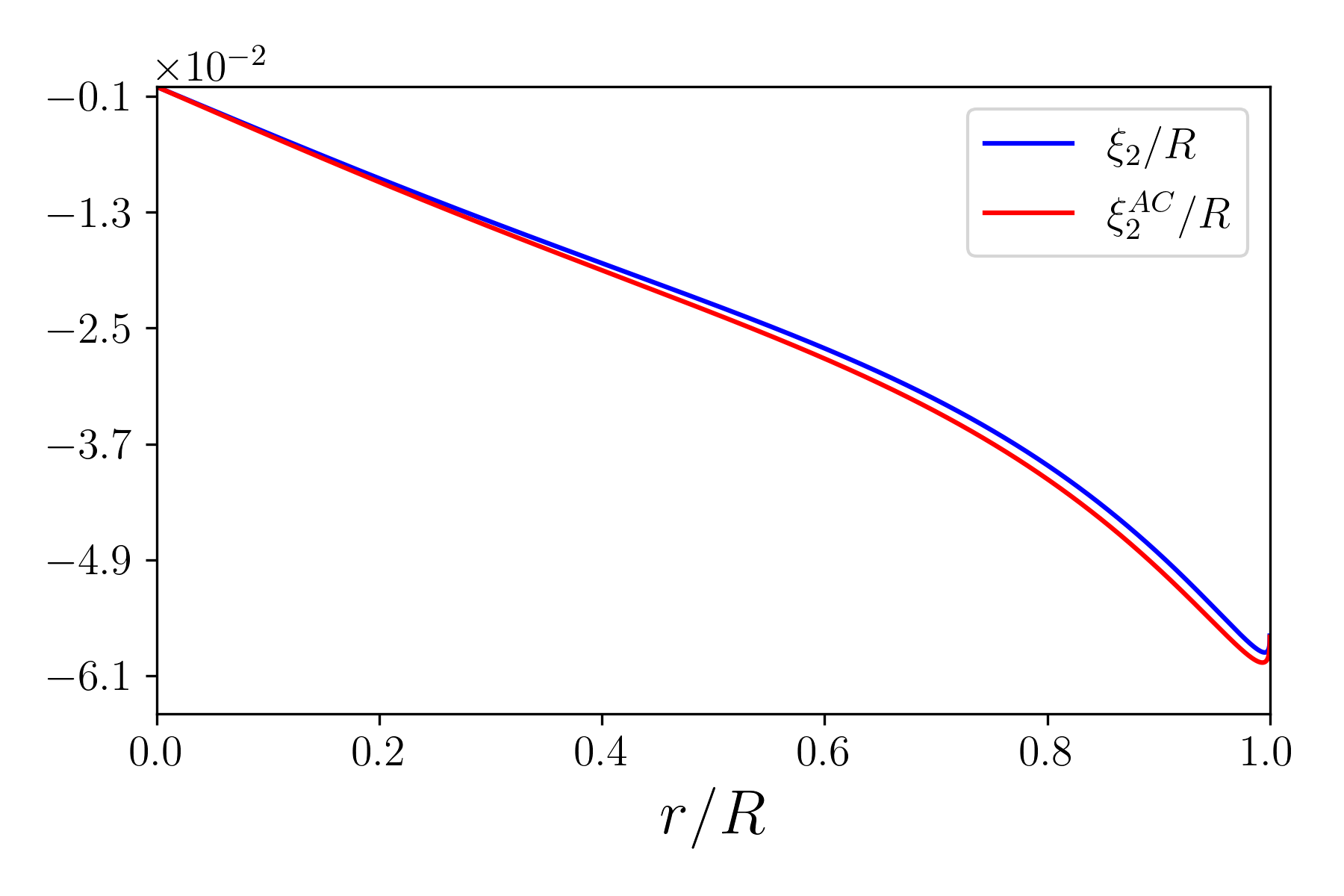}
    \caption{Normalized radial profiles of the deformations $\xi_\ell(r)$ from Eq.~\eqref{eq:xis_r} (blue) and the functions $\xiAC_\ell(r)$ from Eq.~\eqref{eq:xi_l_AC} (red), for $\Delta = 0.5$. The top panel corresponds to $\ell=0$ and the bottom one to $\ell=2$.}
    \label{fig:deformations}
\end{figure}%

\section{Numerical results}
\label{numerical}

We turn now to solve numerically the differential equations of the formalism we present here in order to see the differences arising from the corrections discussed in this article. The results obtained following the amended description will be compared with those in \cite{Andersson_Comer_2001}. We will focus on the comparison of three results: the {contribution to second order to the mass}, $\delta M$, and thus the computation of the total mass, the deformation of the star $\xi_0$ and $\xi_2$, and the Kepler limit $\tOmega_K$.

\subsection{Andersson-Comer model}

In the first place, we solve the background configuration, using the master function provided in \cite{Andersson_Comer_2001}
\begin{align}
    \Lambda_0(n_0^2,p_0^2)=-m_nn_0-\sigma_nn_0^{\beta_n}-m_np_0-\sigma_pp_0^{\beta_p},\label{eq:EoS_AC}
\end{align}
where $m_n$ is the mass of the neutron. As in~\cite{Andersson_Comer_2001} we use the following numerical values
\begin{align}
    \sigma_n=0.2\,m_n,\quad\beta_n=2.3,\quad\sigma_p=2\,m_n,\quad\beta_p=1.95\,.
\end{align}
This master function does not account for entrainment between the two fluids, which translates into $\A_0=\A_0^0=0$. If we impose chemical equilibrium $\mu_\infty=\chi_\infty$ (see the arguments
provided in \cite{Andersson_Comer_2001}) the following relationship between $n_0$ and $p_0$ holds
\begin{equation}
    p_0 = \left(\frac{\beta_n\sigma_n}{\beta_p\sigma_p}n_0^{\beta_n-1}\right)^{1/(\beta_p-1)}.
\end{equation}

\begin{figure}[t]
    \centering
    \includegraphics[width=\linewidth]{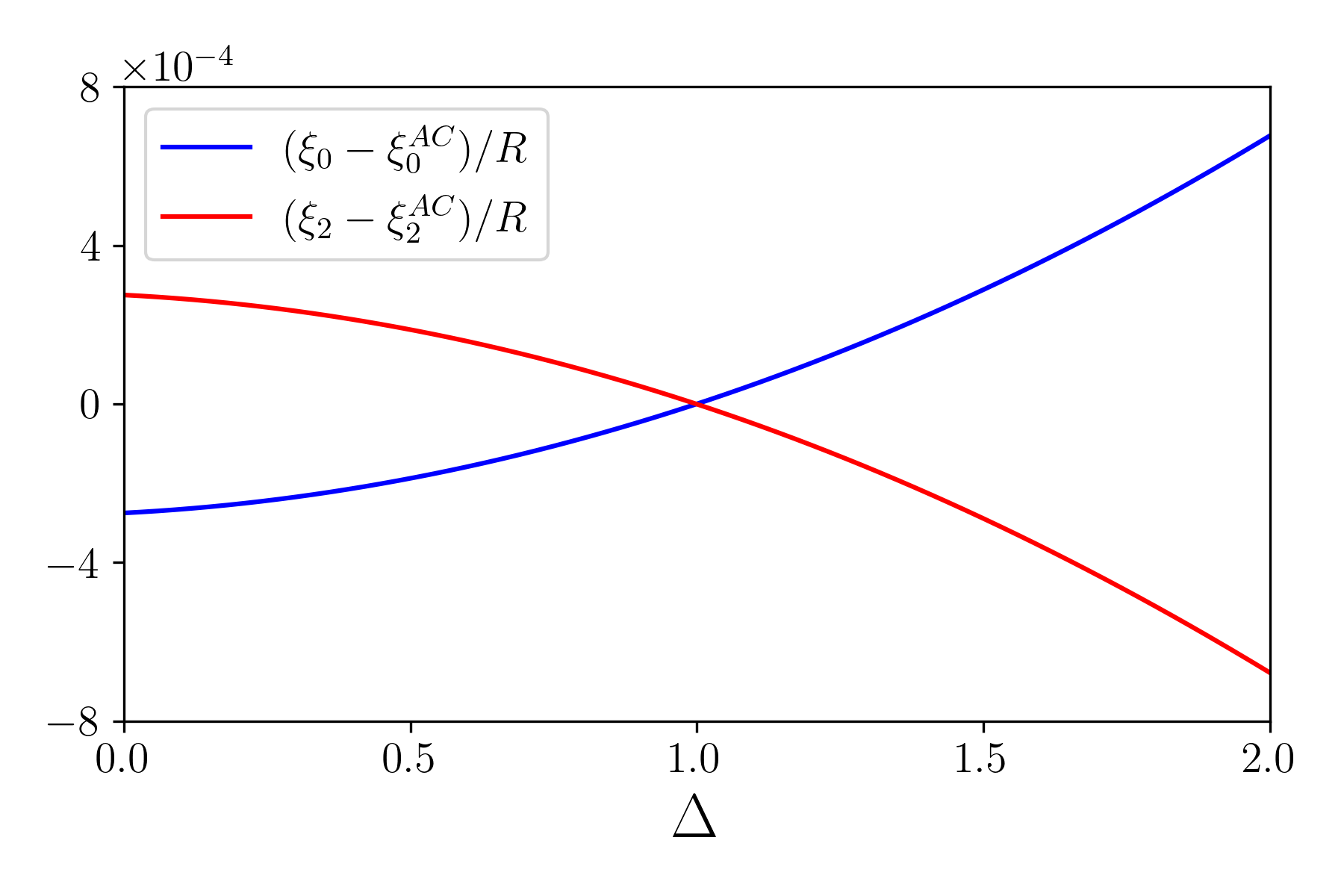}
    \caption{$(\xi_\ell(\rstar) - \xiAC_\ell(\rstar))/\rstar$ for different values of $\Delta$, running from $\Delta=0$ to $\Delta=2$.}
    \label{fig:deformations_boundary}
\end{figure}

For this master function we use units such that $m_n=c=G=1$, and the number densities of protons and neutrons are given in $\si{fm}^{-3}$. We refer to these units as ``code units" (CU). The value of $\Lambda_0$ in the International System of Units (SI) is obtained through the relation
\begin{equation}
    \Lambda_0^{\text{SI}}=\Lambda_0^{\text{CU}}\times c^2\times m_n\times \si{fm}^{-3}.
\end{equation}
It must be noted that this distorts the base units, namely the radial distance $r$, the time $t$ and the mass $m$. To recover the SI units, we need to rescale each variable:
\begin{align}
    &r^{\text{SI}} = r^{\text{CU}}\times c\,\sqrt{\frac{\si{fm}^3}{G\,m_n}},\label{eq:to_meters}\\
    &t^{\text{SI}}= t^{\text{CU}}\times\sqrt{\frac{\si{fm}^{3}}{G\,m_n}},\label{eq:to_seconds}\\
    &m^{\text{SI}} = m^{\text{CU}}\times c^3\,\sqrt{\frac{\si{fm}^3}{G^3\,m_n}}.\label{eq:to_kilograms}
\end{align}

\begin{figure}[t]
    \centering
    \includegraphics[width=\linewidth]{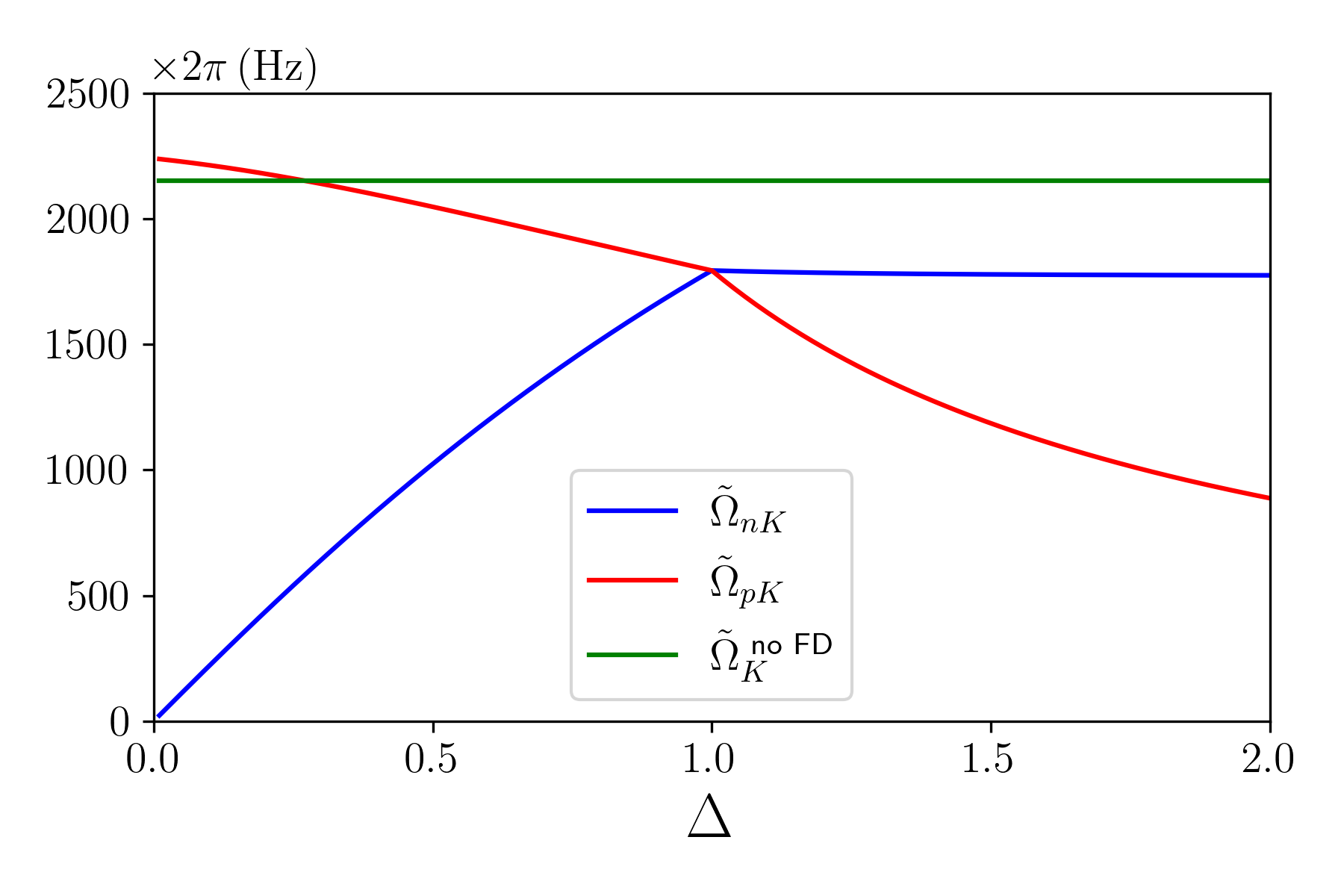}
    \caption{The mass-shedding limits of the neutron and proton fluids (in blue and red) as a function of the relative rotation rate $\Delta$. The Kepler limit corresponds to the largest value $\tOmega_k=\max(\tOmega_n,\tOmega_p)$. The green line corresponds to the Kepler limit with no frame dragging, $\tOmega_k^{\text{no FD}}$.}
    \label{fig:Omega_K}
\end{figure}

Proceeding with the calculation of the background problem, we take $n_0(0)=0.93\,\text{fm}^{-3}$ and solve equations \eqref{eq:dlamb}, \eqref{eq:dnu} and \eqref{eq:first_matrix}, following the procedure of Section \ref{sec:background_comp}. As explained above, to determine the surface of the star we integrate the equations from the center until the pressure reaches its first zero, $\Psi_0(\rstar)=0$. The value $\rstar$ of $r$ for which the pressure first vanishes sets the radius for the background configuration. In accordance with \cite{Andersson_Comer_2001}, the background mass and radius are given by $M=1.41\,M_{\odot}$ and $\rstar=10.08\,\text{km}$, respectively.

As mentioned before, in the original HT formalism the functions describing the metric are assumed to be continuous in the interior of the star, in its exterior, and in the hypersurface separating both regions. However, in the correction of the HT formalism carried out in \cite{ReinaVera2015}, it was shown that $v_0(r)$ presents a discontinuity at the boundary proportional to the energy density at the surface of the star (see Eq.~(\ref{match:funs})). This affects the calculation of $\delta M$ and thus the total mass of the star. For the present two-fluid model, as we can see from Eq.~\eqref{eq:deltaM}, when the number densities of neutrons and protons vanish at the surface of the star, $n_0(\rstar) = p_0(\rstar) = 0$,  which implies
$\press_\ell(\rstar)=0$ by definition, the correction has no numerical effect. As shown in Figure \ref{fig:densities} for the AC numerical model, the number density of protons and neutrons tend to zero at the surface. As a result, in this particular model we
should obtain the same value for the contribution to the mass
at second order as in \cite{Andersson_Comer_2001}. However, we have encountered a discrepancy {in the values.
Explicitly, since $M_T=M+\delta M^{\AC}$ in \cite{Andersson_Comer_2001}, and here $M_T=M+\pert^2 \delta M$
(see \eqref{M_T})
the comparison is given by the quantities
\begin{alignat*}{2}
    &\delta M^{\AC}= 0.091 (\nu_p/ 1\,\si{kHz})^2\,M_\odot,\\
    &\pert^2\delta M=\tOmega_p^2\delta \hat{M}= 0.084 (\nu_p/ 1\,\si{kHz})^2\,M_\odot,
\end{alignat*}
where we have used that $\nu_p$ as defined in \cite{Andersson_Comer_2001}
corresponds to $\tOmega_p/(2\pi)$ here.}
We conjecture this discrepancy might be of numerical origin. A direct comparison with the code by \cite{Andersson_Comer_2001} might shed light on this issue.

\begin{figure}[t]
    \centering
    \includegraphics[width=\linewidth]{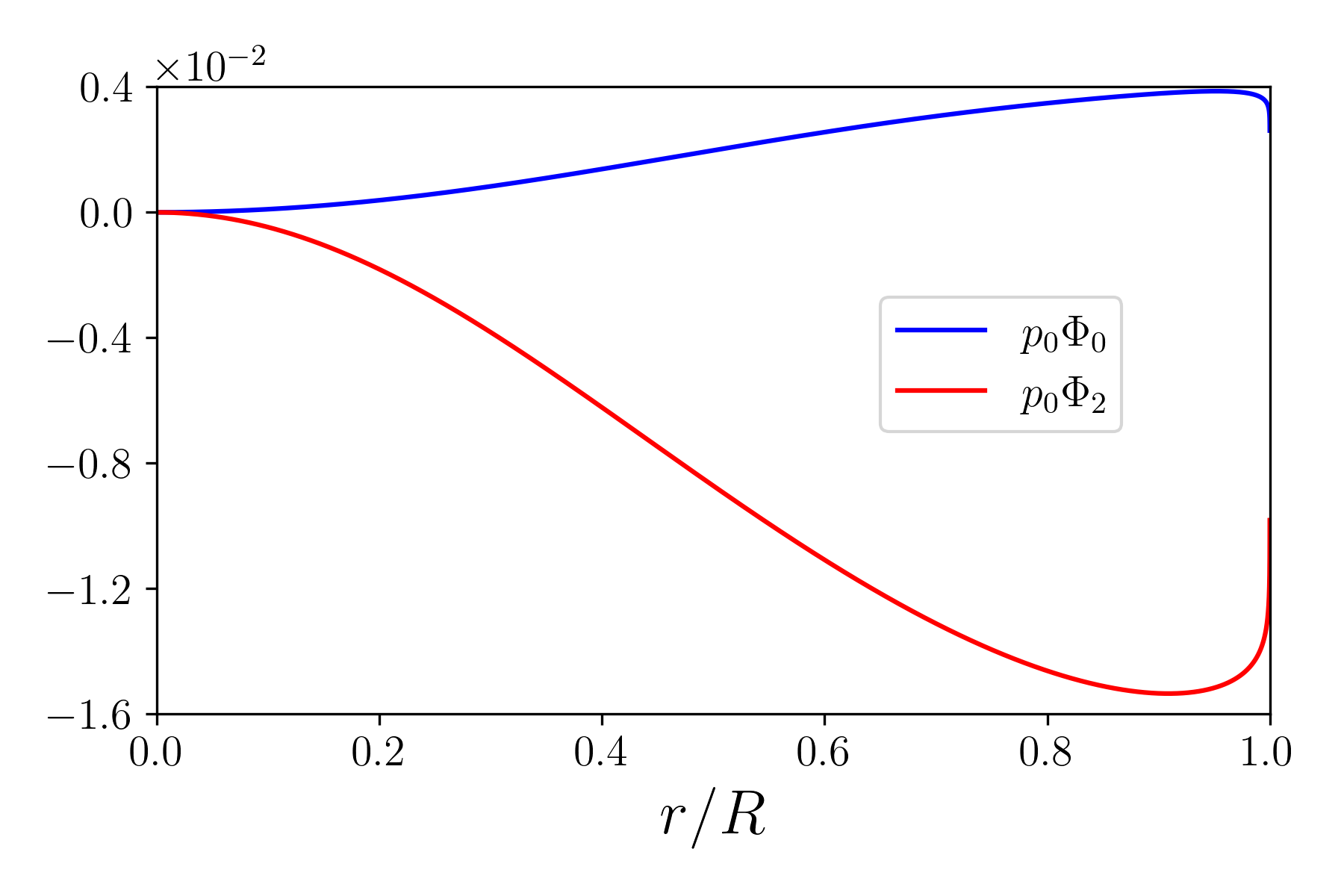}
    \caption{Radial profiles of the rotationally induced change in the proton number density, $p_0\Phi_\ell(r)$, for the AC star model and for $\Delta=1$, in code units. The profiles differ from those in Figure 10 in \cite{Andersson_Comer_2001}.}
    \label{fig:additional_correction}
\end{figure}

On the other hand, in \cite{Andersson_Comer_2001} a very interesting way of calculating the deformation of the star was proposed, that consisted on tracking the surfaces of constant energy density (Eq.~(\ref{eq:xi_l_AC})). However, as  explained in Section \ref{sec:deformations} this analysis has to be carried out using the surfaces of constant pressure (Eq.~(\ref{eq:xis_r})). Concerning the deformation of the star, we compare in Figure \ref{fig:deformations} the functions $\xi_0$ and $\xi_2$ obtained using our formalism (blue curves) with those from the AC formalism \cite{Andersson_Comer_2001} (red curves). As explained in Section \ref{sec:deformations}, for a model without entrainment, $\Omega_n = \Omega_p$,  and in chemical equilibrium, we have $\xi_\ell = \xiAC_\ell$. However, if we set $\Delta\neq1$, we should expect  different results, as Fig.~\ref{fig:deformations} shows for $\Delta=0.5$. The difference between the two results at the surface of the star is plotted in Figure \ref{fig:deformations_boundary} for different values of $\Delta$. Note that due to the steepness of $\xi_\ell^{\AC}(r)$ (and also $\xi_\ell(r)$) near $r=\rstar$, the exact value of $\xi_\ell(\rstar)-\xi_\ell^{\AC}(\rstar)$ will be very sensitive to the precision of the numerical calculations.

The dependence of $\xi_\ell(\rstar) - \xiAC_\ell(\rstar)$ with the perturbation parameter $\pert$ is straightforward, as it scales with $\pert^2$. However, the dependence with $\Delta$ is not so clear from the equations, so we illustrate it in Figure \ref{fig:deformations_boundary}. 
\begin{figure*}[t]
        \includegraphics[width=0.48\textwidth]{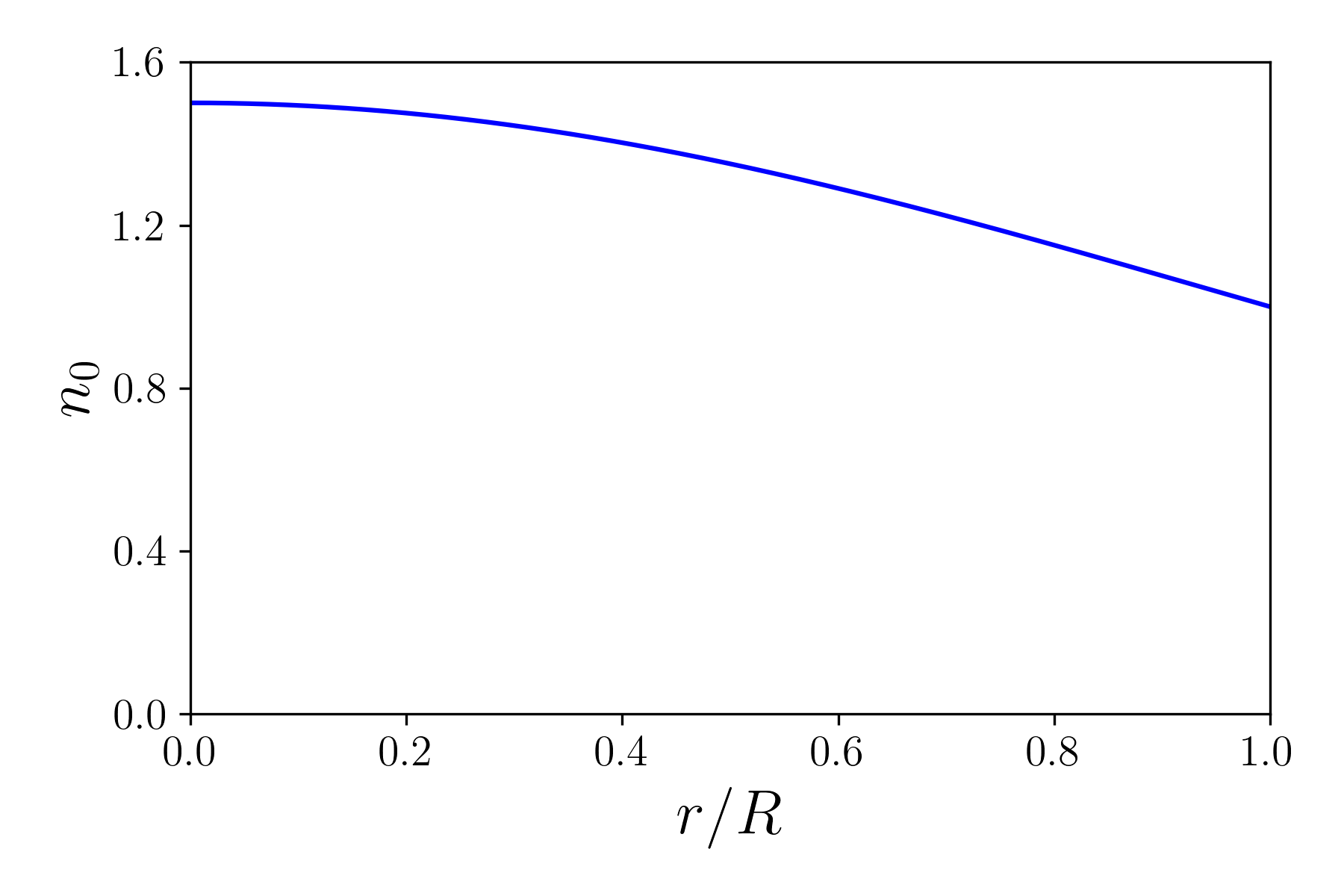}
        \includegraphics[width=0.48\textwidth]{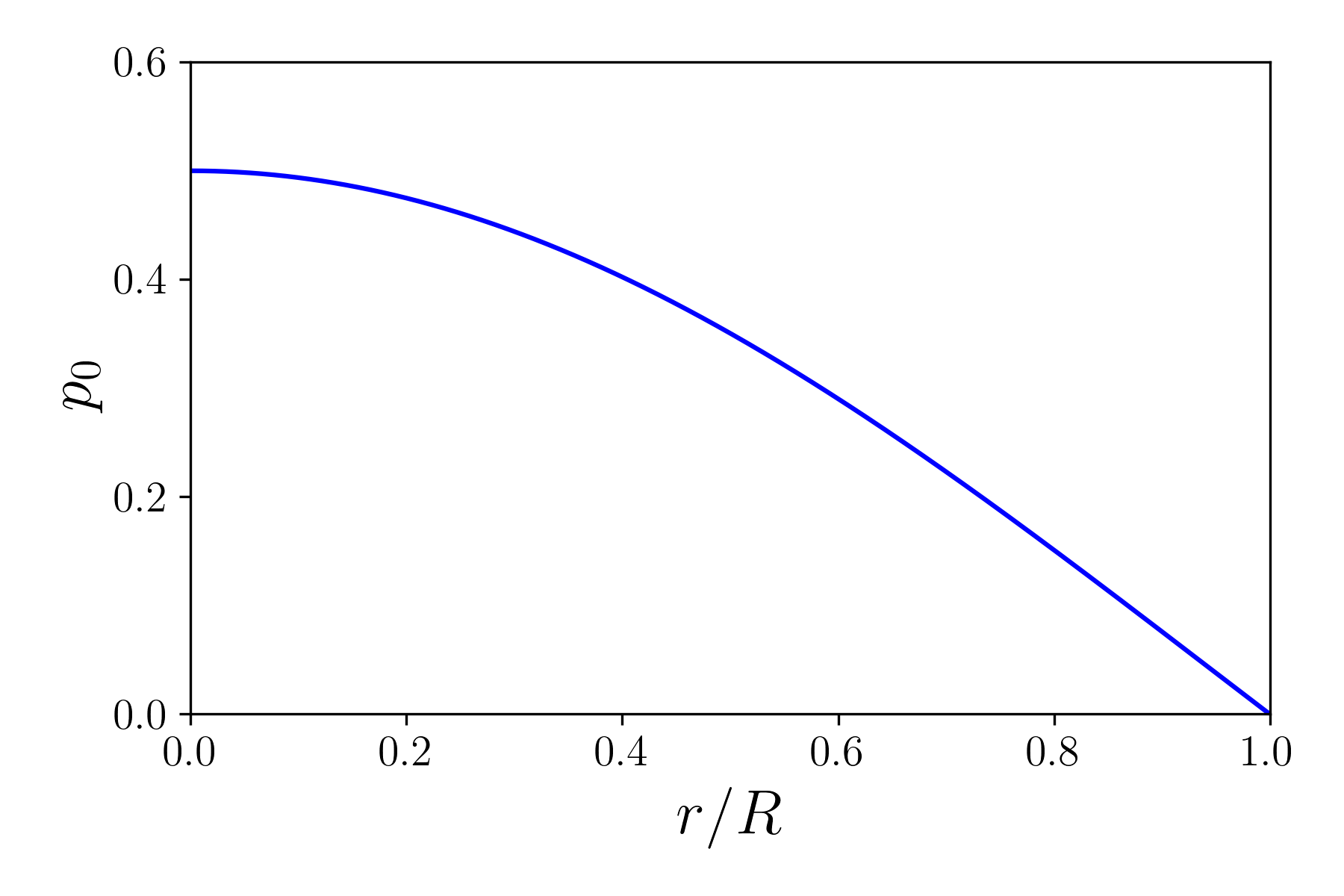}
        \\
        \includegraphics[width=0.48\textwidth]{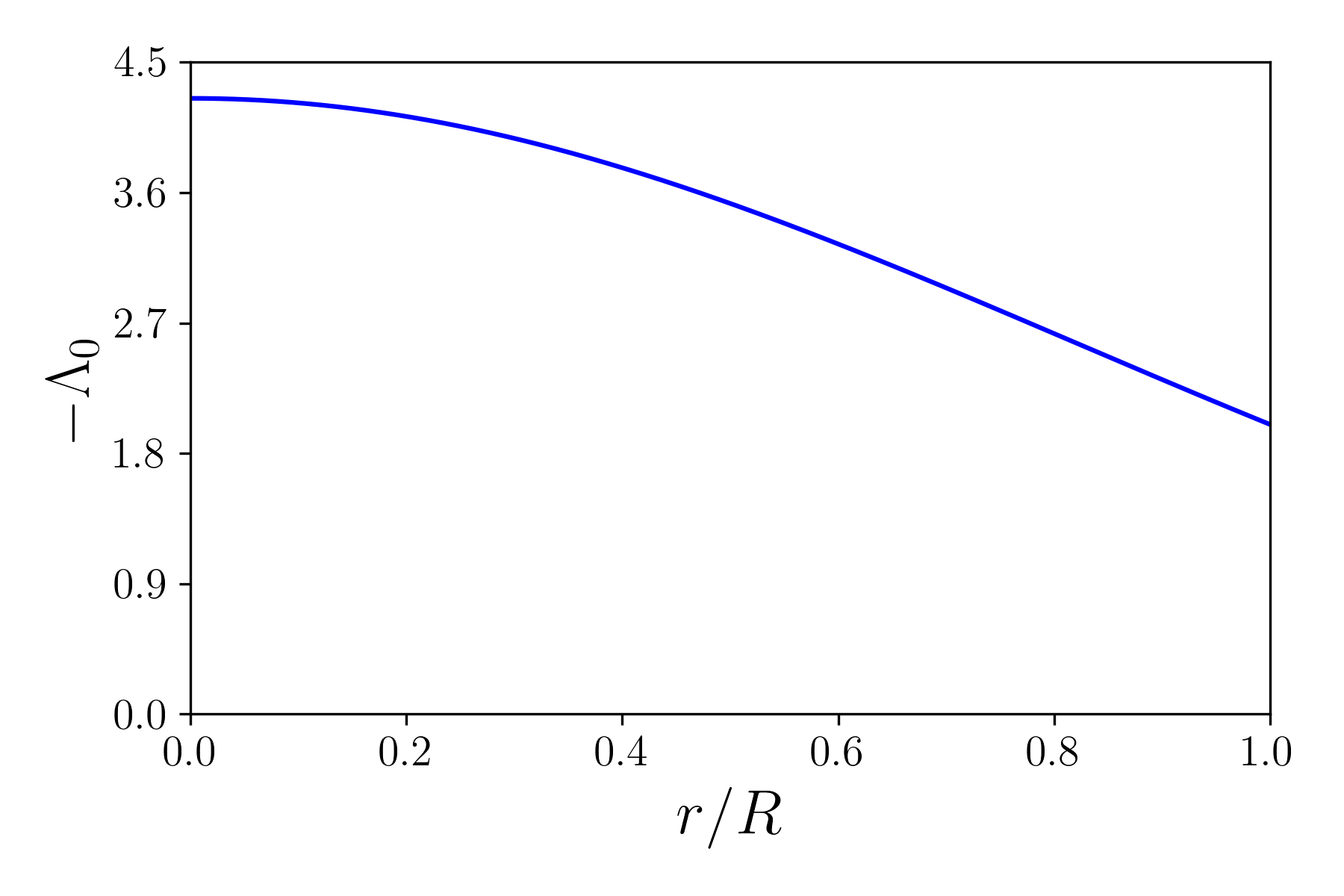}
        \includegraphics[width=0.48\textwidth]{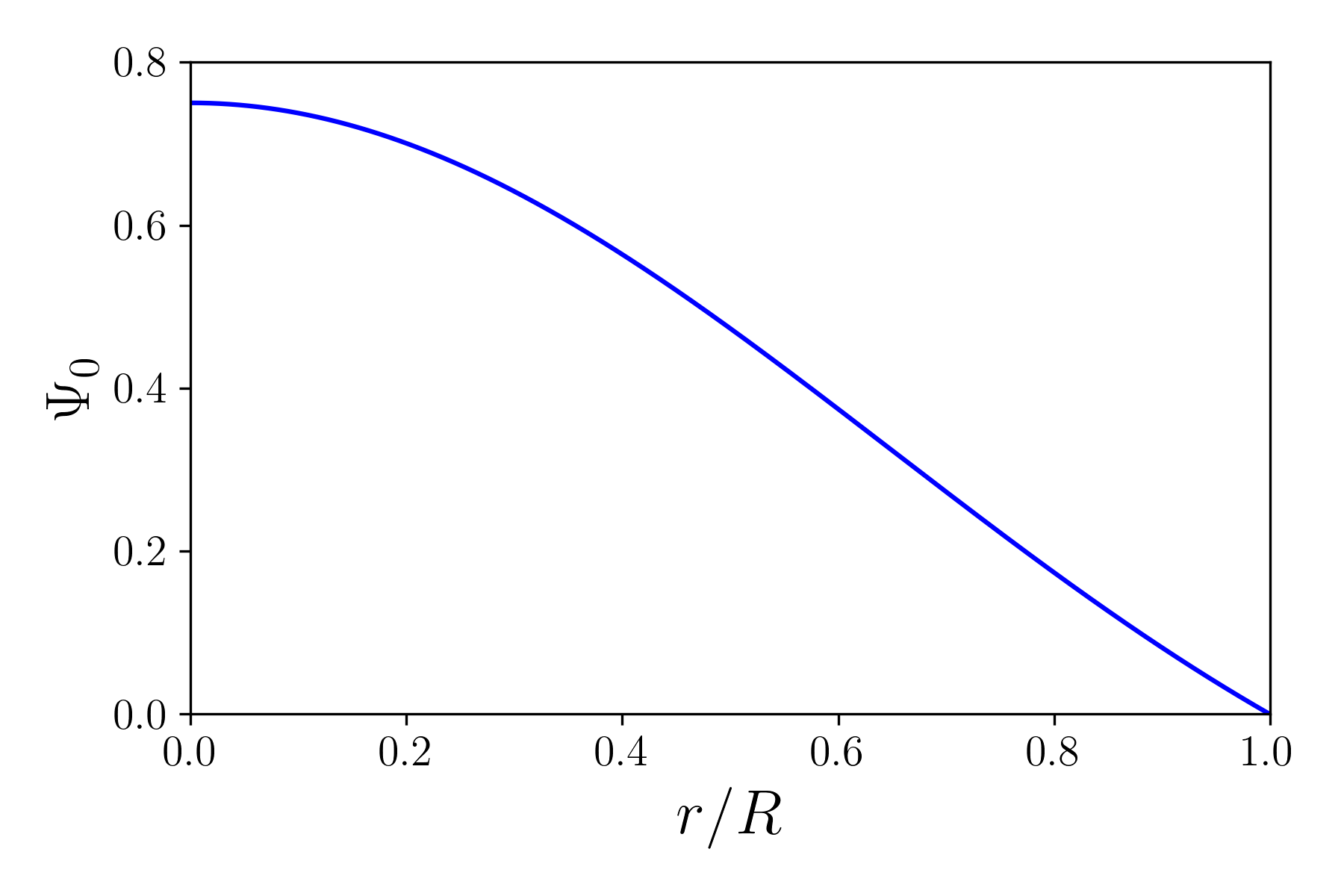}
    \caption{Normalized radial profiles of relevant quantities of the background configuration for the toy model EOS (in code units): number density of neutrons $n_0(r)$ (top-left), number density of protons, $p_0(r)$ (top-right), energy density, $-\Lambda_0(r)$ (bottom-left), and generalized pressure, $\Psi_0(r)$ (bottom right). 
    For this model $n_0(\rstar)\neq0$ and $\Lambda_0(\rstar)\neq0$.} 
    \label{fig:densities_tm}
\end{figure*}

Concerning the computation of the limiting frequency of the star (the Kepler or mass-shedding limit, $\tOmega_K$) we have found a different expression from the one proposed in \cite{Andersson_Comer_2001} (their Eq.~(77)). Moreover, this expression also depends on the deformations of the star $\xi_0$ and $\xi_2$, so the correction of the deformation also implies a different result for $\tOmega_K$, even using the equation reported in~\cite{Andersson_Comer_2001}. The Kepler limits for both neutrons and protons are displayed in Figure \ref{fig:Omega_K} as a function of the relative rotation rate $\Delta$.

As we can see from Eq.~\eqref{eq:OmegaK3}, the mass-shedding limit depends on the frame dragging (FD), i.e.~the distortion of the spacetime manifold as a consequence of the rotation of the star. If we do not consider frame dragging, the Kepler limit frequency will be given by:
\begin{equation}
    \tOmega_K^{\text{no FD}} = \sqrt{\frac{M}{R^3}}.
    \label{eq:noFD}
\end{equation}
As stated in Eq.~(19) of \cite{Andersson_Comer_2001}, the slow-rotation approximation translates into the following inequalities (with $G=1$):
\begin{equation}
    \tOmega_n^2\quad\text{or}\quad\tOmega_p^2\quad\text{or}\quad\tOmega_n\tOmega_p\,\ll \,\frac{M}{R^3}.
\end{equation}

We end this Section by mentioning the discrepancy also found with the AC star model for the radial profiles of the rotationally induced change in the proton number density $p_0\Phi_\ell$, depicted in Figure \ref{fig:additional_correction} (compare with the profiles shown in Fig.~10 of~\cite{Andersson_Comer_2001}).

\subsection{Our toy model}

In order to see the difference of the {contribution
to the mass at second order $\pert^2\delta M$} between the AC formalism and the one reported in this work we propose an EOS which does exhibit a jump in the energy density at the boundary of the star. To do so we consider the following master function 
\begin{equation}
    \Lambda_0(n_0^2, p_0^2, x_0^2) = -(\one n_0 + \two p_0 + \three x_0^2)\,m_n,\label{eq:EoS_TM}
\end{equation}
where $\one$, $\two$ and $\three$ are constants with  dimensions $\dim(\one) = \dim(\two) = 1$ and $\dim(\three) = L^3$. 
This EOS does not attempt to describe any physical system and should be simply regarded as a toy model. Our only purpose in using it is to show the numerical impact of the correction of $\delta M$ discussed in this article. Taking $m_n = 1$ leads to
\begin{alignat}{3}
    &\A_0 = \three, \quad &&\B_0 = \frac{\one}{n_0}, \quad &&\C_0 = \frac{\two}{p_0},\\
    &\A_0^0 = \three, &&\B_0^0=0, &&\C_0^0 = 0.
\end{alignat}
Equations \eqref{eq:first1} and  \eqref{eq:first2} translate into:
\begin{align}
    &n_0^\prime=-\frac{1}{2\three}\,(\two + \three\,n_0)\,\nu^\prime,\\
    &p_0^\prime=-\frac{1}{2\three}\,(\one + \three\,p_0)\,\nu^\prime,
\end{align}
and the generalized pressure \eqref{eq:generalizedpressure} reads
\begin{equation}
    \Psi_0=\three\,x_0^2.
\end{equation}
We impose chemical equilibrium ($\mu_0=\chi_0$) to set the value of $p_0$:
\begin{equation}
    p_0=\frac{\two-\one}{\three}+n_0.
\end{equation}
Both $-\Lambda_0$ and $\Psi_0$ must remain positive along the star, so we take the set $\{\one,\two,\three\}=\{2,1,1\}$ which satisfies the requirements. For the initial value $n_0(0) = 1.5\,\si{fm}^{-3}$ we calculate the profiles of the number densities and pressure of the background configuration. Those are displayed in Figure \ref{fig:densities_tm}.

\begin{figure}[t]
        \includegraphics[width=\linewidth]{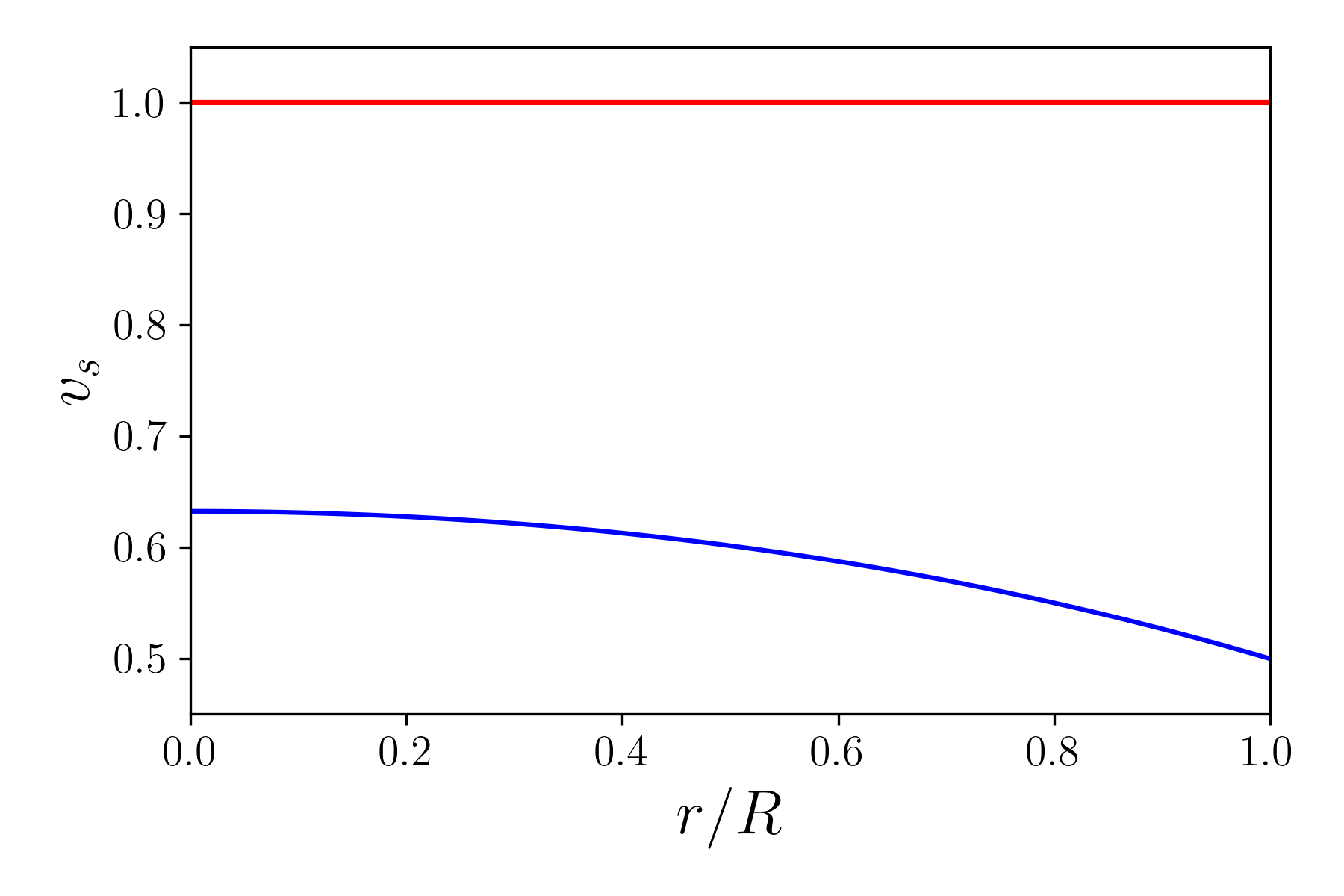}
    \caption{Radial profile of the sound speed for our toy model EOS (blue) and the speed of light (red) in code units. Our EOS preserves causality at all radial points.}
    \label{fig:speed_sound_tm}
\end{figure}

Even though our toy model EOS does not attempt to describe a physical star, we have checked that it is causally consistent with general relativity, i.e.~that the speed of sound is lower than the speed of light at every point of the star. We compute the speed of sound as
\begin{equation}
    v_s^2 = \frac{\partial\Psi_0}{\partial(-\Lambda_0)}=\frac{\partial\Psi_0/\partial r}{\partial(-\Lambda_0)/\partial r}.
\end{equation}
Figure \ref{fig:speed_sound_tm} shows that indeed our EOS preserves causality.

Solving the first and second order ($\ell=0$) system for corotating fluids ($\Delta=1$), we obtain {that the contribution
to the mass at second order is
\begin{equation}
    \pert^2\delta M = 0.0078\, (\nu_p/ 1\,\si{kHz})^2 M_\odot,
\end{equation}
where $\nu_p=\tOmega_p/(2\pi)$, while the formula in \cite{Andersson_Comer_2001} for $\delta M$,
that is, without the correction, leads to
\begin{equation}
    \pert^2\delta M^{\tiny{\mbox{uncorrected}}} = 0.0033\,(\nu_p/ 1\,\si{kHz})^2 M_\odot\,.
\end{equation}
Again, the impact of this correction affects the total mass linearly with $\pert^2$, and thus $\nu_p^2$, but it also depends on the rotation rate between the two fluids.}

\begin{figure}[t]
    \centering
        \includegraphics[width=\linewidth]{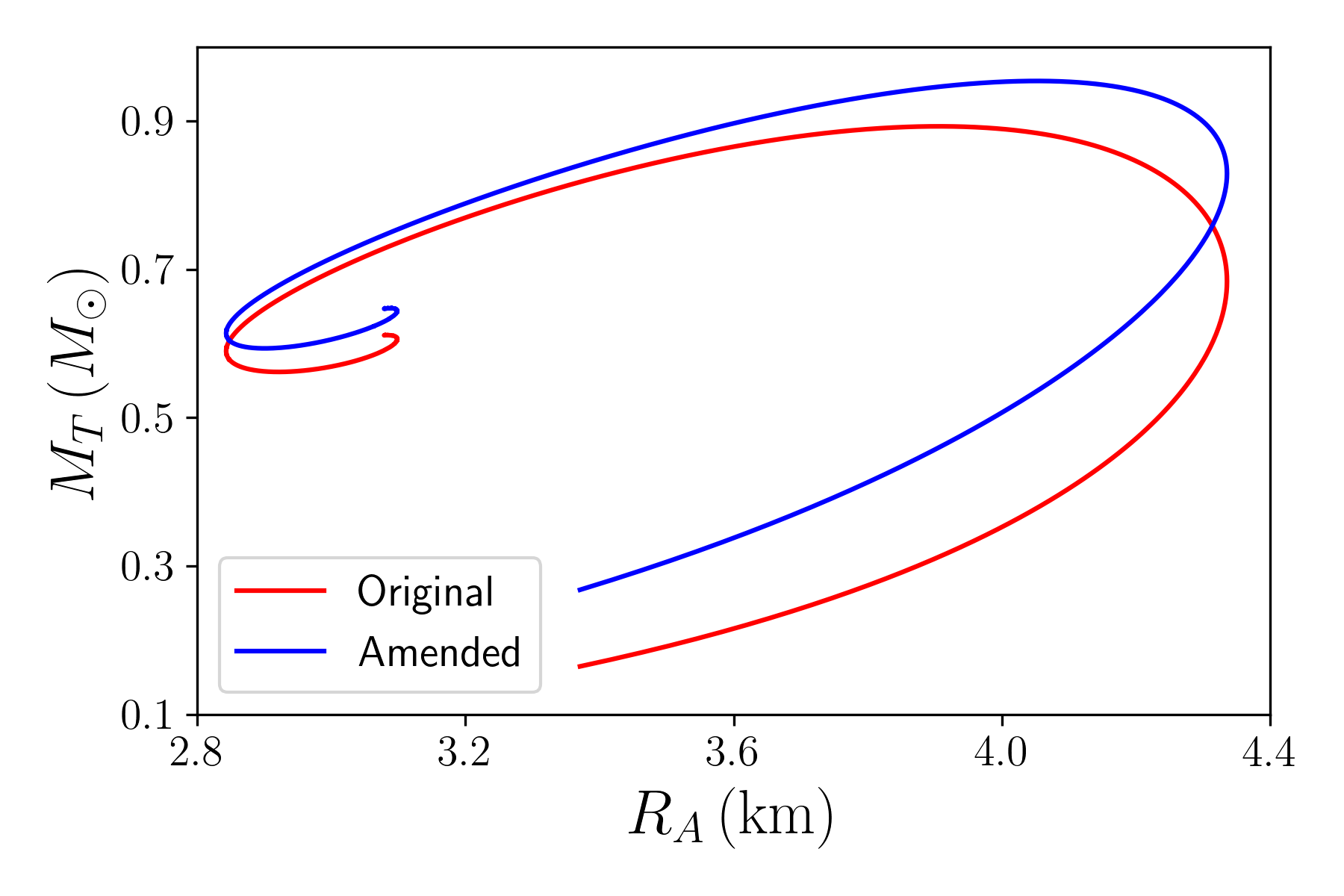}
        \includegraphics[width=\linewidth]{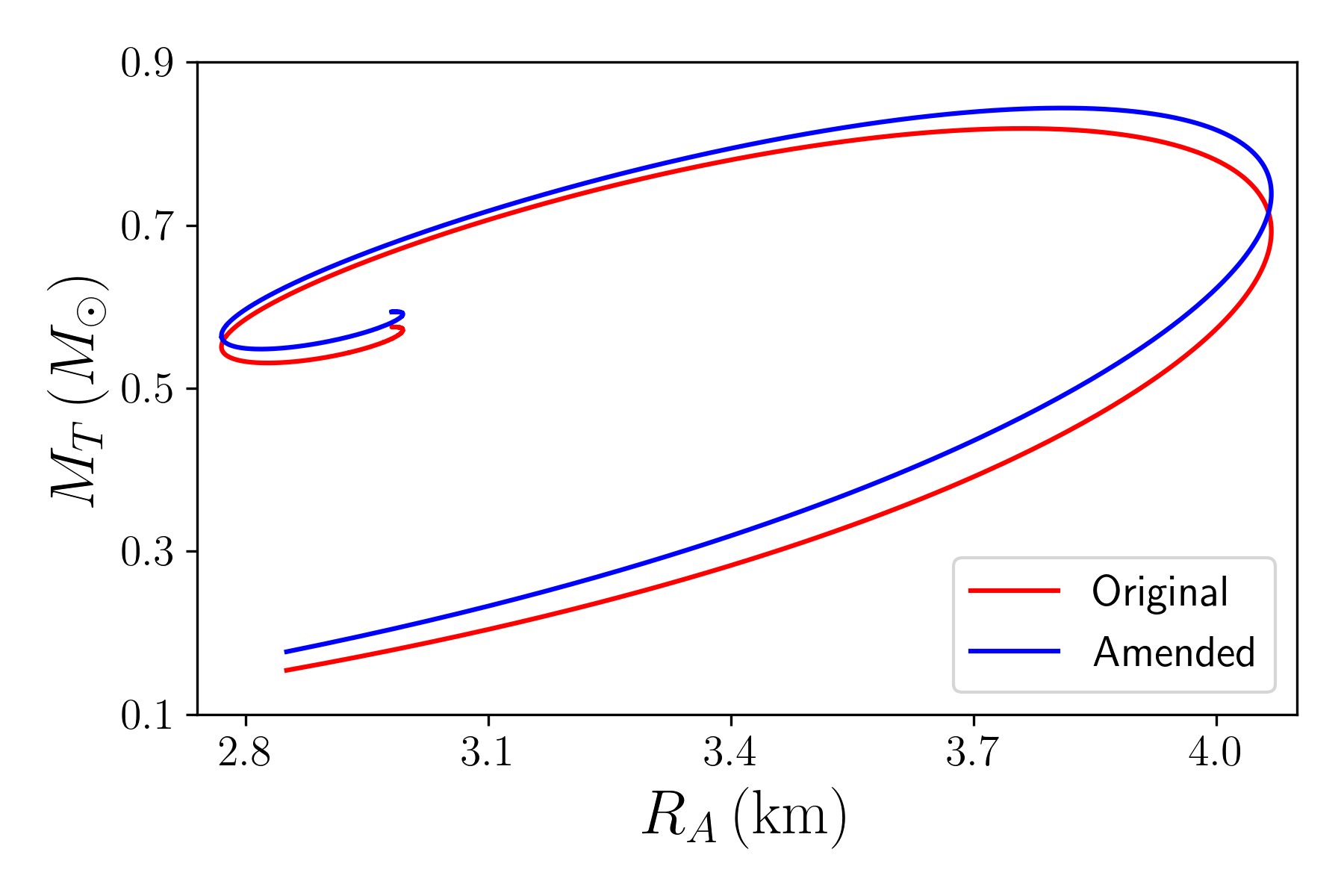}
    \caption{Total mass $M_T$ as a function of the average radius of the star $R_A$ for $\Delta=1$. The top panel corresponds to stars rotating at the Kepler limit without considering frame dragging effects. For the models in the bottom panel the ratio of the polar-to-equatorial radius is set to 0.8. Blue curves  take into account the correction of $\delta M$ in the formalism while red curves do not. To prepare these plots we run the initial value of $n_0$ from $n_0(0)=1.1\,\si{fm}^{-3}$ to $n_0(0)=150\,\si{fm}^{-3}$.}
    \label{fig:deformations2}
\end{figure}

To illustrate the relevance of this correction we numerically build stellar configurations based on our toy model EOS, computing the total mass of the stars as a function of their average radius, $R_A=\rstar + \pert^2\xi_0(\rstar)$. We perform this procedure for two different rotations. 
Our first choice is to consider stars rotating at the Kepler limit without taking into account frame dragging effects (see Eq.~(\ref{eq:noFD})), that is (see Section~\ref{Kepler}),
\begin{equation}
    \pert=\sqrt{\frac{M}{\rstar^3}}\,\frac{1}{\Omega_p}\times\begin{cases}
    1,\quad &\Delta\leq1,\\
    \\
    \displaystyle\frac{1}{\Delta},\quad &\Delta>1.
    \end{cases}
\end{equation}
The top panel of Figure \ref{fig:deformations2} shows the mass of the configurations against the average radius for this case. As expected, the correction in the mass affects in a significant way the total mass the configurations can attain.

In our second choice, we set the rotation of the star so that the ratio between the polar and equatorial radii, $\rstar_P/\rstar_E$, is kept constant, that is,
\begin{align}
    \pert^2=\frac{(\rstar_P/\rstar_E -1)\rstar}{\xi^P - \rstar_P/\rstar_E\,\xi^E}\,,
\end{align}
where $\rstar_{P/E}=\rstar+\pert^2\xi^{P/E}$ with $\xi^P=\xi_0(\rstar)+\xi_2(\rstar)$ and $\xi^E=\xi_0(\rstar)-\xi_2(\rstar)/2$. Mass vs radius diagrams for this second type of configurations are plotted in the bottom panel of Figure \ref{fig:deformations2}. As for the previous case, the differences between the original HT formulation and our revised formalism are quite visible.

\section{Conclusions}
\label{conclusions}

Hartle and Thorne’s {model}~\cite{Hartle1967,Hartle2} provides a perturbative framework to describe the equilibrium configuration of a slowly-rotating isolated compact body in General Relativity. 
In~\cite{ReinaVera2015,ReinaVeranote,reina-sanchis-vera-font2017} it was found that this formalism had to be amended in order to correctly describe {stars with non-vanishing energy density at the surface}. The {amended} version of the formalism yields significant corrections to the total mass of a slowly-rotating relativistic star described by a single-fluid barotropic EOS, which also affects the tidal problem in binary systems (i.e.~the I-Love-Q relations)~\cite{reina-sanchis-vera-font2017}.

In this paper we have discussed slowly-rotating, general relativistic, superfluid neutron stars assuming that the composition of the stars is described by a simple two-fluid model which accounts for superfluid neutrons and all other constituents. Therefore, our work, which has closely followed a previous investigation by Andersson and Comer~\cite{Andersson_Comer_2001}, constitutes an extension of {the} earlier study in the single-fluid case {that amended the HT model}. The backbone of this article has been to put forward the analytical corrections of the HT model applied to the two-fluid problem in general, building on the results from \cite{ReinaVera2015, MRV1, MRV2}. Our specific goal has been to address if the amendment of the computation of the mass of the star found in the HT formalism for the single-fluid case in the presence of discontinuous fields~\cite{ReinaVera2015,ReinaVeranote,reina-sanchis-vera-font2017} also holds in the two-fluid model. We have found that is indeed the case.
{Moreover, we have corrected the determination of the deformation of the star, that in \cite{Andersson_Comer_2001} is obtained using the surfaces of constant energy density following the arguments in \cite{Hartle1967}. The perturbative analysis we have presented provides the deformation, and we have shown how that coincides with the outcome obtained by replacing the surfaces of constant energy density with the surfaces of constant ``pressure''. Let us stress that both procedures coincide in the single-fluid model (with barotropic EOS),
but not in the two-fluid model.}

In order to illustrate the {impact of the corrections to the formalism} we have built numerical stellar models, comparing the {calculation of the total mass} of the star, the deformation of the star, and the Kepler limit of rotation. We have first compared our results with those in \cite{Andersson_Comer_2001} by solving the set of equations for the EOS proposed in the original article along with the same stellar model. Next, we have used a toy-model EOS for which the energy density does not vanish at the boundary of the star to demonstrate that the {corrections} of the HT formalism we {present} in this paper do impact the structure of slowly-rotating superfluid neutron stars.

While the toy model used in this investigation should not be regarded by any means as a realistic description of the composition of actual superfluid neutron stars, it has served the purpose of illustrating the effect of the corrections in the formalism. There are physical situations where large nonzero densities can be attained at the stellar surface, the main example being pure quark EOS (as described by e.g.~the simple MIT bag model~\cite{Zdunik:2000}). Quark matter is self-bound such that low density homogeneous quark matter is unstable with respect to the formation of a dense cluster. Homogeneous nuclear matter shows such an instability, too, depending on the proton fraction. Physically this means that at low densities nuclear clusters and a crust form. To improve the ``realism” of the results reported in this work, a possible extension would require to obtain the master function of~\cite{Langlois:1998,Comer1999} for the case of superconducting quark matter, where a two-fluid situation would arise by assuming pairing only between two flavors (up and down quarks) and have a non-paired third flavor (strange quarks). The master function might be applied in the same way in this case as for the neutron-proton two-fluid model used here, in particular as some models for quark matter resemble technically models applied to the latter case (e.g.~NJL models~\cite{Nambu:1961} are very similar to relativistic density functional models for nuclear matter). Another situation where discontinuities might also affect the computation of stellar equilibrium models within the HT formalism is in the case of superfluid magnetars in which the matter in the core cannot be described with a single fluid approach (since the neutrons are superfluid). In this situation the density of the charged components of the matter EOS (i.e.~protons, electrons and muons) exhibits a jump at the crust-core interface (see e.g.~Figure 1 in~\cite{Gabler:2016}).

Finally, we note that the modifications reported in this work for the total mass of slowly rotating superfluid neutron stars are also present for the tidal problem of a binary system, affecting the I-Love-Q relations of superfluid neutron stars~\cite{Yeung:2021}. We plan to study the tidal problem in a future investigation and the results will be reported elsewhere.

\begin{acknowledgments}
We thank Marc Mars, Micaela Oertel, Michael Gabler, and Nils Andersson for useful comments. Work supported by the Spanish Agencia Estatal de Investigaci\'on (Grants PGC2018-095984-B-I00, PID2021-125485NB-C21, FIS2017-85076-P and PID2021-123226NB-I00 funded by MCIN/AEI/10.13039/501100011033 and ERDF A way of making Europe), the Generalitat Valenciana (Grant PROMETEO/2019/071) and the Basque Government (IT956-16 and IT1628-22). NSG was also supported by the Spanish Ministerio de Universidades, through a María Zambrano grant (ZA21-031) with reference UP2021-044, within the European Union-Next Generation EU. All computations have been performed using open-source computer algebra systems, in particular, SageMath and the free PSL version of REDUCE for the analytical part, and Fortran for the numerical part.
\end{acknowledgments}

\bibliography{super_fluid}
\bibliographystyle{ieeetr}

\end{document}